\documentclass[sigconf]{acmart}
\AtBeginDocument{%
  }

\setcopyright{acmlicensed}
\copyrightyear{2026}
\acmYear{2026}
\setcopyright{cc}
\setcctype{by}
\acmConference[CHI '26]{Proceedings of the 2026 CHI Conference on Human Factors in Computing Systems}{April 13--17, 2026}{Barcelona, Spain}
\acmBooktitle{Proceedings of the 2026 CHI Conference on Human Factors in Computing Systems (CHI '26), April 13--17, 2026, Barcelona, Spain}
\acmPrice{}
\acmDOI{10.1145/3772318.3790823}
\acmISBN{979-8-4007-2278-3/2026/04}

\newcommand{\system}{HOICraft}

\usepackage{ifthen}
\usepackage{algorithm}
\usepackage{algorithmic}
\usepackage{amsmath}
\usepackage{xcolor}
\usepackage{colortbl}
\usepackage{subcaption}
\usepackage{multirow}
\usepackage{makecell}
\usepackage{graphicx}
\usepackage{lscape}
\usepackage{soul}
\usepackage{enumitem}
\usepackage{listings}

\definecolor{baselineBlue}{RGB}{52, 120, 246} 
\definecolor{oursPink}{RGB}{232, 94, 164}     

\usepackage{chi26-554}

\begin{document}

\title{\system{}: In-Situ VLM-based Authoring Tool for Part-Level Hand-Object Interaction Design in VR}

\author{Dohui Lee}
\affiliation{%
  \institution{Graduate School of Metaverse}
  \institution{KAIST}
  \city{Daejeon}
  \country{Republic of Korea}
}
\email{dohui.lee@kaist.ac.kr}

\author{Qi Sun}
\affiliation{%
 \institution{Computer Science and Engineering}
 \institution{New York University}
 \city{New York}
 \country{USA}}
 \email{qisun@nyu.edu}

\author{Sang Ho Yoon}
\affiliation{%
  \institution{Graduate School of Culture Technology}
  \institution{KAIST}
  \city{Daejeon}
  \country{Republic of Korea }}
\email{sangho@kaist.ac.kr }

\renewcommand{\shortauthors}{Dohui Lee, Qi Sun, Sang Ho Yoon}

\begin{abstract}
Hand–Object Interaction~(HOI) is a key interaction component in Virtual Reality~(VR). However, designing HOI still requires manual efforts to decide how object should be selected and manipulated, while also considering user abilities, which leads to time-consuming refinements. We present \system{}, a VLM-based in-situ HOI authoring tool that enables part-level interaction design in VR. Here, \system{} assists designers by recommending interactable elements from 3D objects, customizing HOI design properties, and mapping hand movement with virtual object behavior. We conducted a formative study with three expert VR designers to identify five representative HOI designs to support diverse user experiences. Building upon preference data from 20 participants, we develop an HOI mapping module with in-context learning. In a user study with 12 VR interaction designers, HOI mapping from \system{} significantly reduced trial-and-error iterations compared to manual authoring. Finally, we assessed the usability of \system{}, demonstrating its effectiveness for HOI design in VR.
\end{abstract}


\begin{teaserfigure}
  \includegraphics[width=\textwidth]{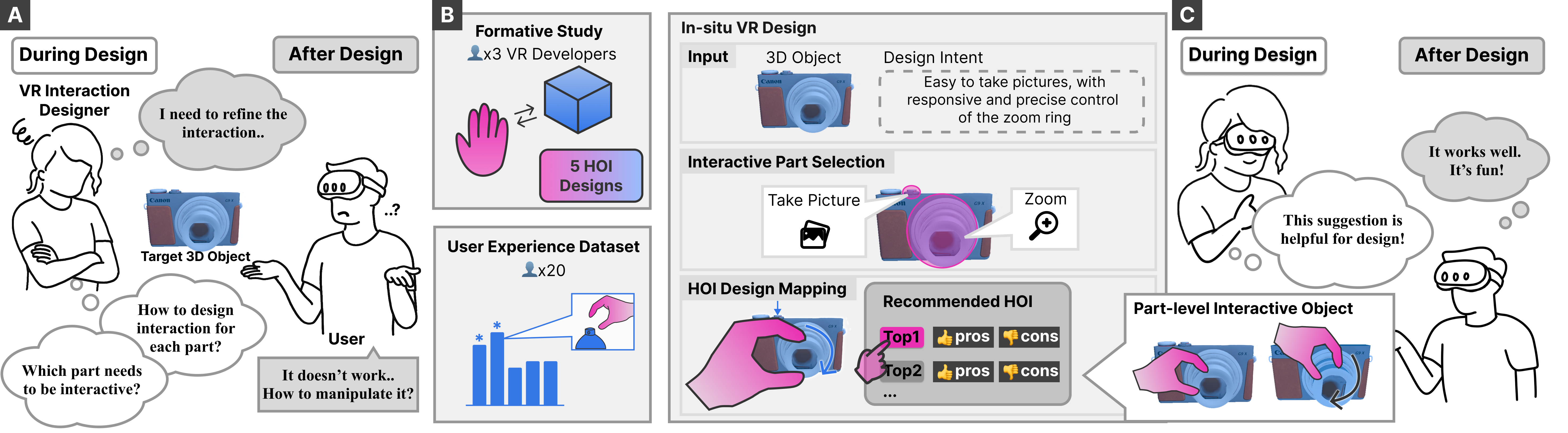}
  \caption{\system{} is an in-situ authoring tool to support part-level Hand-Object Interaction~(HOI) design. (A) During the design process, developers must decide which parts should be interactive and which HOI design to apply. After design, discrepancies often exist between the designers’ intentions and the users' expectations, causing repeated design iterations. (B) We came up with 5 representative HOI designs derived from formative study results and collected user experience data. Our authoring tool enables in-situ VR design for rapid HOI prototyping with part-level detail. (C) By leveraging recommendation-based mapping, we reduce the burden of manual HOI creation and provide a supportive and effective authoring environment.}
  \Description{An illustration divided into three panels explaining the system concept. Panel A (During and After Design) depicts a designer struggling to define interactive parts and a user failing to manipulate the object, highlighting the mismatch between design intent and user experience. Panel B (Method) illustrates the development process, including a formative study with 3 developers, a user experience dataset from 20 participants, and the In-situ VR Design workflow. Panel C (Solution) shows the designer utilizing the system's recommendations, resulting in a helpful suggestion and a successful user experience.}
  \label{fig:teaser}
\end{teaserfigure}

\maketitle
\section{Introduction}
Hand-Object Interaction~(HOI) is an essential component for experiential realism in VR ~\cite{HandProxy, UbiTouch}. Therefore, designing effective HOI in VR is a key element for successful VR interaction design, as overall VR experiences heavily depend on whether users can appropriately manipulate virtual objects~\cite{SelectionManipulation1, SelectionManipulation2}. 
Most VR applications emphasize object-level manipulation, such as grasping and moving whole objects~\cite{3DVirtualObjectManipulation}. Yet, everyday physical interactions are grounded in part-level affordances~\cite{OakInk1,HOI4D}. For example, while a camera is an object, interacting with it involves pressing the shutter button to capture an image or rotating the zoom ring to adjust magnification. This highlights that enabling fine-grained part-level interactions is essential for achieving natural HOI in VR.

To support part-level interaction, prior work has largely concentrated on the mechanics of object motion---such as mesh cutting and motion constraint definition in immersive authoring~\cite{AffordIt} or transforming 3D meshes into articulated objects~\cite{ArticulateAnymesh}. While these methods improve defining how parts can move, they provide limited functionality guidance on how those parts should interact according to the user's action---namely, selection and manipulation~\cite{SelectionManipulation1, SelectionManipulation2}. To better understand this challenge, it is important to consider how selection and manipulation are realized in practice. These components can be designed at different levels of fidelity, which directly influence the user experience. For instance, physics-based manipulation can provide highly realistic interactions~\cite{3DVirtualObjectManipulation,HoloDesk,EfficientPhysicsHOI}, whereas gesture-based manipulation often emphasizes convenience~\cite{virtualgrasp,HandInterface}. Similarly, object interactions may involve direct~\cite{argelaguet2013survey, sorli2021fine, mendes2016benefits} or animation-based manipulations~\cite{GesturAR, AffordIt}. Accordingly, the design process becomes considerably more complex at the part-level, because each part requires a different combination of selection and manipulation depending on its properties. In practice, such decisions are often made based on the designer's prior experience and intuition, resulting in high effort and inconsistent outcomes. Furthermore, designers must repeatedly determine which HOI design best aligns with their intent, making the process time-consuming. To streamline this process, in-situ authoring tools have been leveraged by providing immediate feedback~\cite{coelho_authoring_2022}. However, while these immersive environments facilitate evaluation, the configuration of interaction mechanics itself remains manual and step-by-step. This underscores the need for intelligent support in part-level HOI design.

To this end, advancements in foundational models have opened up new opportunities for supporting designers. Recent studies have introduced LLM-powered authoring tools to assist in design workflows, such as prototyping~\cite{SonifyAR,Epigraphics} and creativity support~\cite{AIdeation,LumiMood, Amuse}. Moreover, Vision–Language Model~(VLM) has begun to support XR authoring by interpreting visual context and object semantics~\cite{dogan2024augmented, hou2025echoladder}. In this context, extending such AI-assisted authoring to HOI design is appealing, as it can help designers select part-level HOI designs that better reflect both object properties and user needs, going beyond manual trial-and-error. Currently, these foundation-model–based systems largely rely on the general knowledge, making them well-suited for creative ideation. Yet, for HOI design, which must be grounded in real-world user behaviors, such knowledge can yield suggestions that appear plausible but fail in practice. For instance, LLMs frequently recommend physics-based HOI as the most ``realistic'' option across parts and contexts, even though alternative interactions may provide a more natural fit. Such mismatches highlight a critical limitation in supporting user-dependent HOI design.

To fill these gaps, we introduce \system, a VLM-based in-situ authoring tool designed to support the part-level HOI design in VR. By interpreting the designer's intent together with object properties extracted by the VLM, \system{} identifies which parts should become interactive. In addition, \system{} leverages LLMs with in-context learning based on user preference data to support determining suitable HOI designs for each part. Furthermore, \system{} supports designers to customize HOI designs by providing detailed properties. With all features integrated into the in-situ environment, designers can rapidly map, test, and refine HOI with reduced manual effort, ultimately supporting more effective and purposeful VR experiences. 

We designed and evaluated \system{} through three studies. First, a formative study with three experts in commercial VR content development identified practical requirements for HOI design. This study revealed five representative HOI design criteria that cover diverse user experience levels and contexts. In particular, we observed that their approaches to selection (e.g., physics-, gesture-, and contact-based) and manipulation methods (e.g., direct manipulation and animation) varied considerably depending on the target user experience and context. Second, a comparative study with 20 participants of varying VR experience levels examined these five criteria. We collected both quantitative and qualitative preference data, which informed the development of our HOI mapping module. This module leverages an in-context learning framework to automatically suggest suitable HOI mappings for interactive VR parts based on general user preferences. Finally, we developed \system, an immersive AI-driven HOI authoring tool that enables designers to efficiently create, test, and refine part-level HOI designs directly in VR. To assess its effectiveness, we conducted a study with 12 VR designers, comparing manual HOI mapping with our HOI mapping module, and a usability test for the immersive design workflow and authoring tool.

Our main contributions are as follows:
\begin{enumerate}
    \item We propose \system{}, an AI-assisted in-situ authoring tool that enables designers to create, test, and refine part-level HOI in VR with reduced manual effort.
    \item We identify five representative HOI design criteria with diverse selection and manipulation strategies, derived from a formative study with professional VR developers.
    \item We develop a user-preference–driven HOI recommendation module that leverages in-context learning on empirical user data, supporting AI-assisted co-design by suggesting practical HOI mappings for virtual objects.
\end{enumerate}

\section{Related Work}
\subsection{Hand-Object Interaction in VR}
HOI refers to how humans use their hands to directly manipulate objects, and it represents a primary way humans take action in the world. Recent works have attempted to generate HOI in 3D to advance VR applications~\cite{HOIGPT, HOIDiffusion, Text2HOI, ReconstructingHOI}. However, these efforts primarily focus on synthesizing realistic HOI motions or visualizations, and often require additional coding or post-processing to translate them into interactive behaviors. To support interactive VR experiences, HOI design can be understood along two fundamental dimensions: (1) how the hand selects object parts (selection), and (2) how the selected part subsequently responds (manipulation)~\cite{SelectionManipulation1, SelectionManipulation2}. 

Prior work on physics-based interaction typically attaches colliders to both the hand and virtual objects to support selection and manipulation through collision~\cite{3DVirtualObjectManipulation,HoloDesk,EfficientPhysicsHOI}. The key factor in these approaches is collision handling through simulating the virtual object’s behavior according to physics laws~\cite{PhysicsHOIPrecision}. Another line of research has explored gesture-based interaction, using grasping gestures~\cite{virtualgrasp} or gesture imitation of objects~\cite{HandInterface} to retrieve and manipulate them, as well as mapping user actions to customized object behaviors~\cite{GesturAR}. While these approaches enrich the design space, they do not address how different user groups may require distinct interaction strategies. To bridge this gap, our study identifies five representative HOI patterns through a formative study that considered a broad range of users. These representative interactions serve as a foundational design layer in our authoring tool, enabling designers to configure HOI according to their intended goals and target audiences.

\subsection{AI-assisted Intelligent Workflow Support in XR}
AI models have the potential to reduce workflow burdens by interpreting abstract design intents to provide tailored support. To this end, prior work has explored automating content creation through text-to-scene generation~\cite{LLMR,PromptGenAIVRScene,SynthesizingPlayReadyVrSCenes, Text2VRScene, dang2023worldsmith, vachha2025dreamcrafter} and text-to-interaction through code synthesis~\cite{DreamCodeVR, GROMIT, ConnectVR, kurai2025magicitem}. At the same time, others have kept designers in the loop, where LLMs support creativity by suggesting design alternatives~\cite{yuan2022wordcraft,AIdeation}, providing inspiration~\cite{gero2022sparks,choi2024creativeconnect,LumiMood,Amuse}, and accelerating iterative prototyping to reduce trial-and-error costs~\cite{SonifyAR,Epigraphics,VRCopilot}.

In XR, recent works have begun to integrate such AI-assisted workflows into intelligent authoring systems. Previous works alleviate manual effort by supporting the creation of interactive simulations~\cite{gunturu2024augmented, suzuki2020realitysketch} or by enabling programmable logic for physical objects~\cite{monteiro2023teachable}. Building on these intelligent authoring approaches, more recent systems incorporate LLMs to assist prototyping~\cite{VRCopilot, SonifyAR} and interpret multimodal inputs as authoring assistants~\cite{hu2025gesprompt, wang2025can}. Researchers also employed VLMs to better understand visual context and object semantics, helping systems provide more knowledgeable authoring support~\cite{dogan2024augmented, hou2025echoladder}. Collectively, these advancements align with the broader vision of ``programmable reality,'' where AI and XR technologies converge to make the physical environment increasingly configurable and interactive~\cite{suzuki2025programmable}. In this evolving context, extending AI-assisted authoring to HOI design is appealing, as designers also require intelligent support to configure how users interact with these virtual parts in ways that match their design intent.

However, relying solely on the general knowledge embedded in current foundation models presents a challenge. HOI design requires grounding in real-world user experiences, as people perceive and value interactions differently depending on the context and user abilities. Consequently, without such empirical grounding, relying only on LLM knowledge risks generating interactions that look plausible but fail in practice. To address this, we collect user experience data on HOI designs and integrate it directly into the recommendation process through an in-context learning approach~\cite{dong2022survey}, which has shown competitive performance without parameter updates~\cite{dai2022can}. This enables the system to adapt its recommendations to better reflect actual user performance and needs.

\subsection{In-situ Authoring in VR Applications} 
In VR, in-situ authoring tools offer a major advantage by enabling a ``What you see is what you get'' environment for content creation~\cite{coelho_authoring_2021,coelho_collaborative_2019}. Such tools provide an intuitive interface providing immediate feedback during the design process, reducing both interaction steps and time required for creation~\cite{coelho_authoring_2022}. 

Building on these advantages, prior work has explored diverse applications of in-situ authoring, such as animation~\cite{kurai2025magicitem, vogel2018animationvr}, 3D CAD modeling~\cite{arslan2025tinkerxr}, and defining interactive behaviors. Several studies have demonstrated that users can create, edit, and evaluate 3D content directly in an in-situ environment~\cite{mine1995isaac,ens2017ivy}. In addition, researchers integrated visual programming into in-situ workflows, enabling objects to react dynamically to user input~\cite{FlowMatic, hedlund2023blocklyvr, eroglu2024choose, wang2020capturar, li2025intereconMemorable, sung2024hapticpilot}. Similarly, Programming by Demonstration~(PbD) approaches have been employed to support interaction authoring~\cite{GesturAR, chauvergne2023authoring, sayara2023gesturecanvas}.

Still, prior systems have largely focused on defining objects' interactive behaviors at a coarse level (e.g., user or object). Moreover, despite the benefits of in-situ feedback, these approaches often require manual, step-by-step specification of behaviors, which becomes increasingly labor-intensive for part-level interactions. What remains underexplored is a fine-grained interaction authoring method to support part-level selection and manipulation. Here, our work introduces an authoring framework for part-level HOI mapping that captures designer intent and recommends suitable interactive parts and HOI mappings. By integrating the intelligent suggestions with immediate feedback, the proposed in-situ authoring makes less labor-intensive part-level HOI design.

\section{Formative Study: Identifying Representative HOI Design in VR}
\label{sec:3FormativeStudy}
We conducted interviews with professionals to investigate how part-level HOI is currently designed in practice. Our findings revealed key challenges in the process, underlying design space, and key metrics influencing HOI design.

\subsection{Participants}
We recruited three professional VR developers (all male; aged 26 to 35) with varying VR interaction development experience (2–6 years, $M=4$, $SD=1.63$) as participants. All participants were actively engaged in creating VR applications, using the Unity engine as their primary development environment. Table~\ref{tab:participants} summarizes the demographics of the participants. The developers who participated in the IRB-approved study received compensation for their time with 50 USD.

\begin{table}[b]
\centering
\caption{Demographics and experience information of formative study participants.}
\begin{tabular}{cccccl}
\toprule
ID & Age & Gender & Experience & Programming skills \\
\midrule
P1 & 29 & Male & 4 years & Unity engine, Blender \\
P2 & 35 & Male & 6 years & Unity engine, Unreal engine \\
P3 & 26 & Male & 2 years & Unity engine \\
\bottomrule
\end{tabular}
\label{tab:participants}
\Description{Demographics of participants in the formative study, including age, gender, years of VR experience, and programming skills.}
\end{table}

\subsection{Interview Process and Analysis}
\label{sec:FormativeInterviewProcess}
Each participant engaged in a semi-structured interview lasting 2 hours. The study consisted of four stages: \textit{Introduction, Pre-Interview, HOI Design Task \& Follow-up Questions, and Post-Interview}. During the \textit{Introduction}, we explained the study purpose and procedure and obtained consent to record the interview. For \textit{Pre-Interview}, we explored typical workflows and challenges in designing part-level HOI. We showed participants a 3D padlock mesh featuring both rotation and translation mechanics, alongside its interactive version (unlocked with a virtual hand) and asked them to describe the process they would follow. In the \textit{HOI Design Task}, participants were asked to select an interactive part within a given object and design HOI for that part across multiple VR contexts (e.g., game, social, and training). We observed their decision-making process and asked follow-up questions to better understand their reasoning. During the \textit{Post-Interview}, we asked participants to describe the factors they considered, and the trade-offs they encountered in their HOI designs. All interviews were recorded and transcribed for analysis. We employed affinity diagramming~\cite{AffinityDiagram} to extract key themes and identify developers’ needs in designing part-level HOI in VR.

\subsection{Findings}
\subsubsection{Challenges in Part-Level HOI Creation in VR and Design Implications}
From our formative study, we identified 3 key challenges in creating part-level HOI in VR.

\textbf{Part-Level Decomposition~(Challenge~1).} Participants noted that determining the appropriate interactive part for design intention is not straightforward. They often adjust decomposition levels depending on the contextual information and repeatedly make subjective decisions. Adjusting decomposition levels often leads to inconsistency and cognitive burden. As P1 explained, \textit{``Complexity tends to follow a pattern by context, but we often have to customize it differently depending on the case''}. Similarly, P2 noted, \textit{``It’s hard to decide which parts to allow interaction with—it’s different for every object, so it's difficult to generalize''}. P3 also admitted, \textit{``As I kept working, I realized my approach to part decomposition wasn’t very consistent''}. To address this, we introduced the \textit{Part Prioritizer}, which leverages intent text to automatically select the most relevant parts for interaction. By prioritizing these parts as a starting point, the system ensures greater consistency across projects and reduces the cognitive burden of manually deciding decomposition levels.

\textbf{Designing HOI for Diverse Contexts~(Challenge~2).} Participants noted that interactions that seemed appropriate during design often failed in practice, requiring substantial redesign to accommodate actual user abilities and contexts. P1 mentioned, \textit{``It often turns out that the more realistic the interaction, the harder it is for users to operate. So we redesign it differently''} and P2 noted \textit{``We implemented physical forces to match real-world mechanics, but users couldn’t operate it properly—so we created a version that adjusts automatically''}. To support this, we designed \textit{HOI Mapper} based on user preference data to improve alignment with user expectations and reduce the risk of repeated redesign.

\textbf{Lack of Tool Support for Iterative HOI Refinement~(Challenge~3).} Designing part-level HOI often requires iterative refinement, as developers must repeatedly adjust and test how each part responds to actual user interaction. These trial-and-error cycles are essential to ensure the interaction feels natural and usable in context, but they can be highly time-consuming. Despite this, existing tools provide limited support for rapidly prototyping and evaluating these interactions. P2 noted, \textit{``What takes a lot of time is the trial and error—figuring out whether people can actually do it, and whether this approach even works. We keep adjusting things over and over, and that eats up a lot of time''}. To address this, we adopted in-situ environment that provides immediate feedback. This can accelerate testing cycles, reduce trial-and-error attempts, and lower the overall overhead of iterative refinement.

\subsubsection{Observed Patterns for VR HOI Design}
\label{sec:ObservedHOIPatterns}
We identified recurring patterns of HOI design, with diverse approaches to selection and manipulation depending on the design context.

\textbf{Selection.} Participants described several techniques for how the virtual hand selects a part, which generally fell into 2 main approaches: physics-based selection and snapping-based selection. With physics-based selection, colliders on each finger joint simulate realistic grasping~(P1-3). This approach was sometimes used to deliberately increase difficulty, such as stacking tasks~(P1), but it often caused accidental movements, making it unsuitable for standard tasks~(P1-2). Snapping-based selection includes two patterns: automatic selection upon contact~(\textit{``Like a doorknob that grabs itself when the hand just gets close''~(P2)} and \textit{``Sometimes we make it snap easily without fixing the exact pose''~(P1)}) and gesture-based selection~(\textit{``For handles like drawers or doorknobs, people naturally try to grab them, so fixed gestures worked well''~(P1))}.

\textbf{Manipulation.} After selection, participants also differed in how the part responds to user input. Some participants highlighted real-time manipulation closely coupled with hand motion, \textit{``We wanted the part to move exactly with the hand—like pulling a lever''~(P3)}. In contrast, another participant emphasized pre-defined animations as a way to ease manipulation for novices, \textit{``In VR, inexperienced users sometimes struggle even to grab and open a door. For such cases, we design interactions where merely touching the doorknob triggers the action''~(P2)}.

\subsection{Deriving an HOI Design Space}
\label{sec:4HOIDesign}
\begin{figure*}[t]
\centering
   \includegraphics[width=0.8\textwidth]{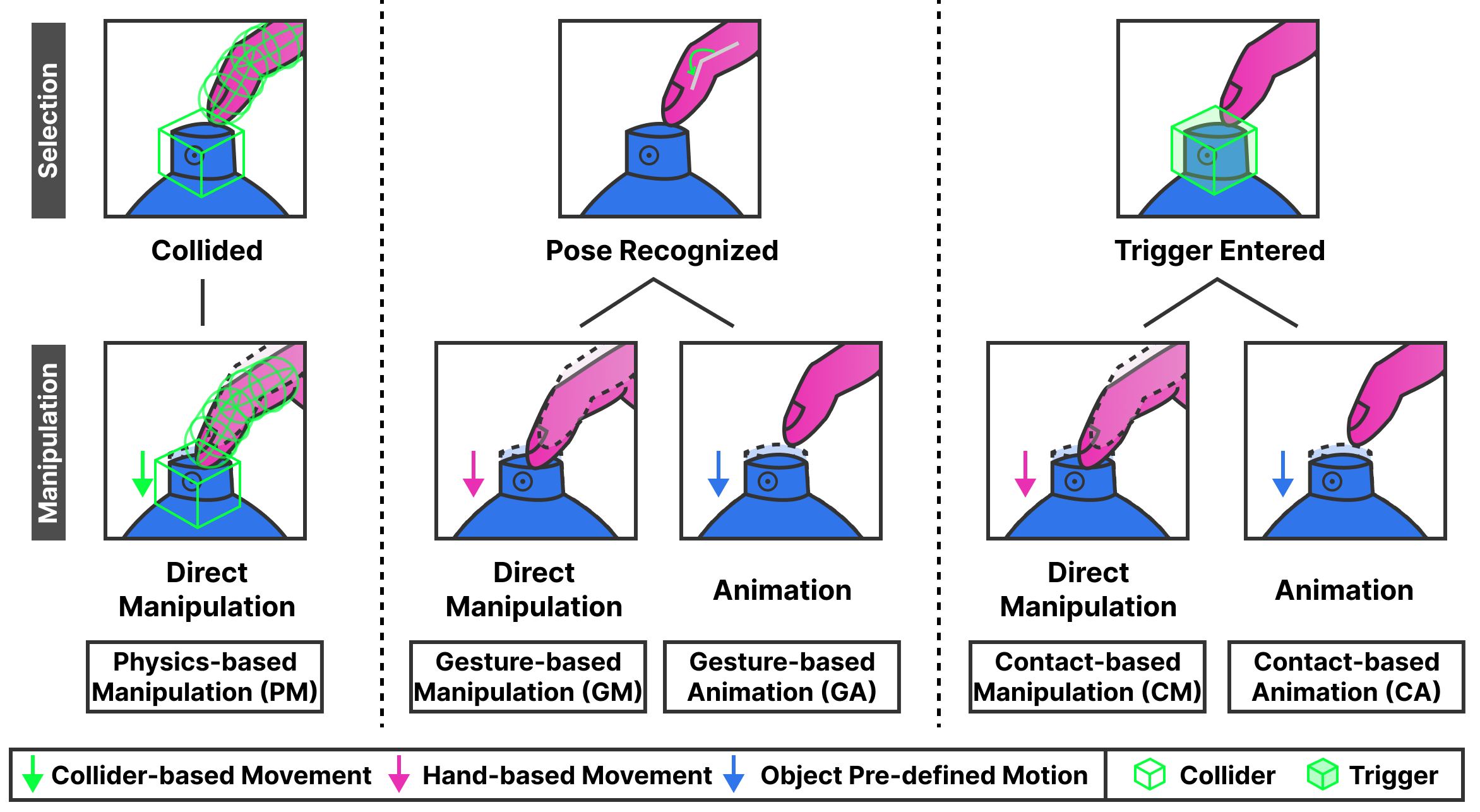}
   \hfil
\caption{Representative HOI design methods for virtual objects from the formative study. \textbf{PM} refers to physics-based manipulation. \textbf{GM} and \textbf{GA} represent gestured-based manipulation and animation accordingly. \textbf{CM} and \textbf{CA} mean contact-based manipulation and animation.}
\Description{A diagram illustrating the HOI design space divided into three sections. The left section shows Physics-based Manipulation (PM), where a virtual hand collides with a green collider to move the object. The center section shows Gesture-based methods: Gesture-based Manipulation (GM) where the object follows hand movement, and Gesture-based Animation (GA) where the object plays a predefined motion upon selection. The right section shows Contact-based methods triggered by a green box: Contact-based Manipulation (CM) where the object follows the hand, and Contact-based Animation (CA) where it plays a predefined motion.}
\label{fig:HOIDesign}
\end{figure*}

Based on observed patterns from Section~\ref{sec:ObservedHOIPatterns}, we derived a design space for representative HOIs along two key dimensions: selection method (collider-, gesture-, and contact-based) and manipulation response (direct manipulation and animation). While these dimensions yield six combinations, we exclude the physics–animation case, since physics-based selection is inherently collider-driven and produces physical behaviors that conflict with predefined animations. Our design space consists of five representative HOI methods as shown in Figure~\ref{fig:HOIDesign}. 

\textbf{Physics-based Manipulation~(PM).}
Interaction occurs through colliders attached to both the hand and the object, relying on physics simulation for selection and manipulation. For example, the user must physically move the hand to the opposite side of a hinged door's handle to apply a pulling force. Light contact pushes the object away via collision forces.

\textbf{Gesture-based Manipulation~(GM).} 
Users acquire an object by performing a specific gesture. Once acquired, the object continuously follows the hand within a predefined motion range while the gesture is maintained, and it is released when the gesture breaks, a threshold is crossed, or tracking is lost.

\textbf{Gesture-based Animation~(GA).} 
Same gesture-based acquisition as \textbf{GM}, but the object plays a predefined animation~(e.g., opening/closing the door) instead of continuous hand mapping. Each animation is triggered once per gesture cycle for clarity and consistency.

\textbf{Contact-based Manipulation~(CM).} 
Objects are acquired by entering the trigger region, without requiring a gesture. The object then follows the hand within a predefined motion range until the hand exits the trigger. Unlike \textbf{PM}, the object is kinematically driven by the hand upon contact, behaving as if magnetically attached. For instance, touching a door handle instantly couples the door to the hand, causing it to mirror the hand's trajectory exactly, regardless of pulling force.

\textbf{Contact-based Animation~(CA).} 
Same contact-based acquisition as \textbf{CM}, but the object responds with a predefined animation upon entry, as in \textbf{GA}. Animations are triggered only once per out–in cycle, preventing unintended repetition. This ensures that playback reflects deliberate user intent.

\subsection{Key Metrics Influencing HOI Design}
\label{sec:keymetrics}

\begin{figure}[b]
\centering
   \includegraphics[width=\linewidth]{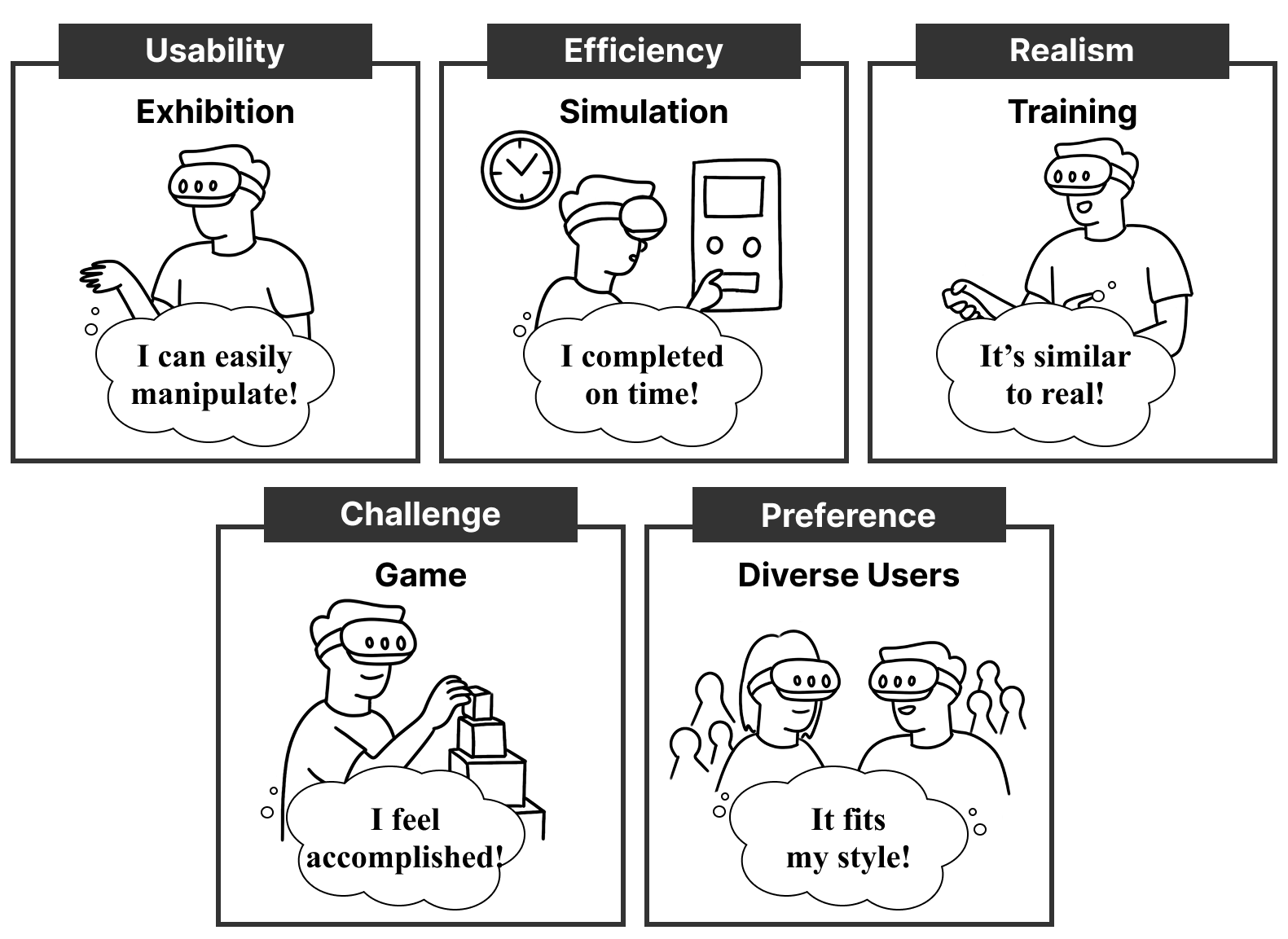}
   \hfil
\caption{Key metrics for HOI design. The metrics reflect diverse design intents of different user types and contexts.}
\Description{Five illustrations representing different evaluation metrics. The first panel (Usability – Exhibition) shows a person gesturing and saying "I can easily manipulate!" The second (Efficiency – Simulation) shows a person operating a panel saying "I completed on time!" The third (Realism – Training) shows a person with a fire extinguisher saying "It’s similar to real!" The fourth (Challenge – Game) shows a person stacking blocks saying "I feel accomplished!" The fifth (Preference – Diverse Users) depicts diverse users saying "It fits my style!"}
\label{fig:KeyMetrics}
\end{figure}

Our formative study revealed that VR developers made their HOI design decisions considering diverse contexts. From these observations, we identified four main design metrics (Figure~\ref{fig:KeyMetrics}) that shaped the decision: \textit{Usability, Efficiency, Realism,} and \textit{Challenge}. In addition to these developer-driven factors, we also add user \textit{Preference} as an additional metric since it serves as a base guideline when designers do not have a clear design intent.

\textbf{\textit{Usability}.} Participants emphasized that interactions should feel smooth and comfortable, particularly for novice users or exhibition settings where realistic manipulation is often impractical. In these cases, they simplified the design so that actions could be triggered with minimal effort—for example, selecting without explicit gestures (P1, P2) or automatically triggering when the hand touched the object (P2). As one participant explained, \textit{``We don’t always replicate real-world interactions exactly. Sometimes realism makes manipulation harder''~(P1)}.

\textbf{\textit{Efficiency}.} Participants emphasized that users should be able to complete the task in a timely and reliable manner. \textit{``Even if the interaction feels less realistic, we usually change it so that users can still perform the task. It's more important that it works, even if the experience is slightly worse''~(P2)}. This was particularly critical in contexts such as exhibitions or training applications (e.g., VR factory safety simulations), where ensuring reliable task completion was prioritized.
    
\textbf{\textit{Realism}.} In training contexts, realism was seen as essential, requiring accurate reproduction of real-world actions: \textit{``In training scenarios, we tried to consider all possible realistic operations and movements that a user would perform with the real object''~(P2).} \textit{``The goal in designing for training was to closely replicate the real-world functionality of the object''~(P3)}. This aligns with prior work that highlights realism as a key dimension in VR design~\cite{stanney2003usability,bowman2012questioning,kapralos2014overview,ragan2015effects}.

\textbf{\textit{Challenge}.} In gaming, experienced VR users often preferred more demanding interactions, so participants deliberately raised the interaction threshold: \textit{``Experienced VR users tend to prefer more diverse interactions''~(P1). } \textit{``In VR games, users are generally familiar with the environment, so even if we set the interaction threshold a bit higher, they can handle it well''~(P2).} We refer to this factor as \textit{Challenge}, following the terminology used in the Game Experience Questionnaire ~\cite{ijsselsteijn2013game} which emphasizes not only difficulty but also the accompanying sense of effort, learning, and accomplishment.

\textbf{\textit{Preference}.} Beyond these developer-driven factors, we also considered users’ direct preferences for different HOI designs. This factor reflects the user’s subjective evaluation of how well HOI aligns with their expectations and comfort. We treated preference as a key metric because it can serve as a guideline in cases where developers may not have a clear design intention, providing insight into which HOI designs users find most acceptable.

\section{Study 1: Empirical Data Collection of HOI Design Key Metrics}
\label{sec:5DataCollection}
We conducted a user study on the representative HOI designs~(Sec~\ref{sec:4HOIDesign}) to examine how developer-identified key metrics~(Sec~\ref{sec:keymetrics}) align with users’ actual perceptions at the part level. This empirical data informed the formulation of the HOI mapping module (Sec~\ref{sec:hoimapping}). The study comprised two main tasks with distinct objectives.

\textbf{Task 1 (User Experience Comparison)}. Participants provided both \textit{Preference} rankings and Likert-scale ratings (\textit{Usability} and \textit{Realism}) for the HOI designs. This task captured users' subjective perceptions of how different designs were experienced.
    
\textbf{Task 2 (Performance Comparison)}. Participants performed a direct manipulation task to assess \textit{Efficiency} (i.e., timely and reliable task completion). Because this task required fine control and adjustment, it also revealed participants' perception of \textit{Challenge}. Accordingly, we compared three selection methods combined with direct manipulation (\textbf{PM, GM, CM}), while excluding animated manipulations (\textbf{GA, CA}) whose fixed motions limit the control and adjustment.

Through these tasks, we were able to identify how different HOI designs corresponded to each key metric across various object-part pairs.

\subsection{Object-Part Dataset}
\begin{figure}[t]
\centering
   \includegraphics[width=\linewidth]{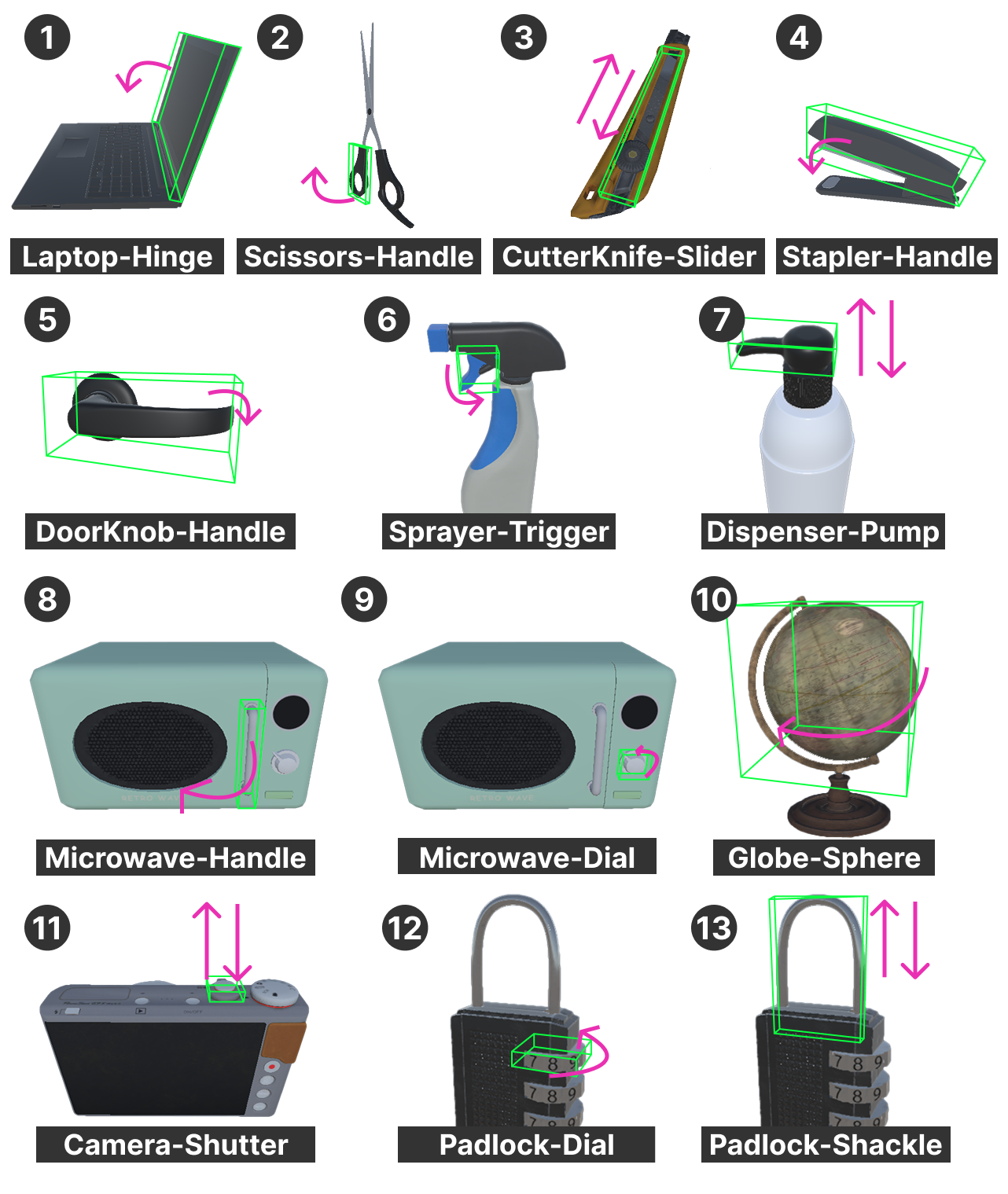}
   \hfil
\caption{Selected 13 object–part pairs for user experience data collection in Study 1. Green bounding boxes show collider/trigger regions defined based on the mesh size of each part. Pink arrows indicate embedded motion constraints, which we predefined according to the natural behavior of each object (straight arrows for translation and curved arrows for rotation). }
\Description{A set of 13 3D objects, each with an interactive part marked by a green bounding box and pink motion arrows. Objects include a laptop (hinge), scissors (handle), cutter knife (slider), stapler (handle), door knob (handle), sprayer bottle (trigger), dispenser bottle (pump), microwave (handle and dial), globe (sphere), camera (shutter button), and two padlocks (dial and shackle). Straight arrows indicate translation (e.g., pressing, sliding), and curved arrows indicate rotation (e.g., turning, spinning).}
\label{fig:Study1PartSet}
\end{figure}

To construct a representative object–part dataset, we referred to HOI datasets in both computer vision~\cite{ARCTIC, HOI4D, AffordPose, ContactDB, YCB, OakInk1} and VR~\cite{HandInterface, virtualgrasp, UbiTouch}. We selected 20 everyday objects covering diverse characteristics (e.g., affordances and size), while prioritizing part-level interaction over static or deformable ones. We then generated object–part candidates based on expert decomposition results in \textit{HOI Design Task} (Sec~\ref{sec:FormativeInterviewProcess}) and refined them to fit the study duration, excluding overly similar interactions (e.g., pulling the trigger of a sprayer vs.\ a drill). A total of 13 object-part pairs are selected as shown in Figure~\ref{fig:Study1PartSet}.

\subsection{Implementation}
\label{sec:Implementation}
The study was developed using Unity 2022.3.43f1 and ran on a Meta Quest 3 HMD connected to a PC (Intel i9-13900K CPU, RTX 4070 Ti GPU) via a Link cable. We used the Meta Interaction SDK\footnote{\url{https://developers.meta.com/horizon/documentation/unity/unity-isdk-interaction-sdk-overview}} and Meta Hand Tracking API\footnote{\url{https://developers.meta.com/horizon/documentation/unity/unity-handtracking-overview}} to obtain 3D joint poses and hand state information. We implemented all HOI designs on top of the aforementioned APIs, Unity’s built-in PhysX engine for physics-based behaviors, and kinematic mapping for gesture- and contact-based variants.

\subsection{Study Setup}
We recruited 20 participants (8 female), aged 19 to 33 ($M=25.15$, $SD=3.91$). Based on a demographics survey, 7 participants were identified as beginners in terms of VR experience (3 with no prior experience and 4 with less than 1 hour), 9 participants as intermediate users (1–10 hours), and 4 participants as advanced users (one with 10–100 hours and three with more than 100 hours of experience). Regarding hand-tracking experience, 10 participants reported no prior experience, while 10 participants reported having some experience. In terms of handedness, 19 participants reported being right-handed and 1 participant reported being left-handed. The study was approved by the IRB and participants received compensation for their time with 30 USD. 

The study consisted of three phases: (1) \textit{introduction and training}, (2) the \textit{user experience comparison task} using a think-aloud protocol to capture ranking reasoning, and (3) \textit{the performance comparison task} followed by a semi-structured interview. The study lasted for 2 hours in total.

\subsubsection{Introduction and Training Session}
We collected demographic information and introduced the five HOI designs to ensure participants understood how each worked. A training session followed using a practice object (not included in the main tasks) configured with all HOI designs side by side, allowing participants to directly compare them. Training continued until participants reported they could reliably distinguish among them, usually within five minutes.

\subsubsection{Task 1: User Experience Comparison}
\begin{figure}[!t]
\centering
   \includegraphics[width=\linewidth]{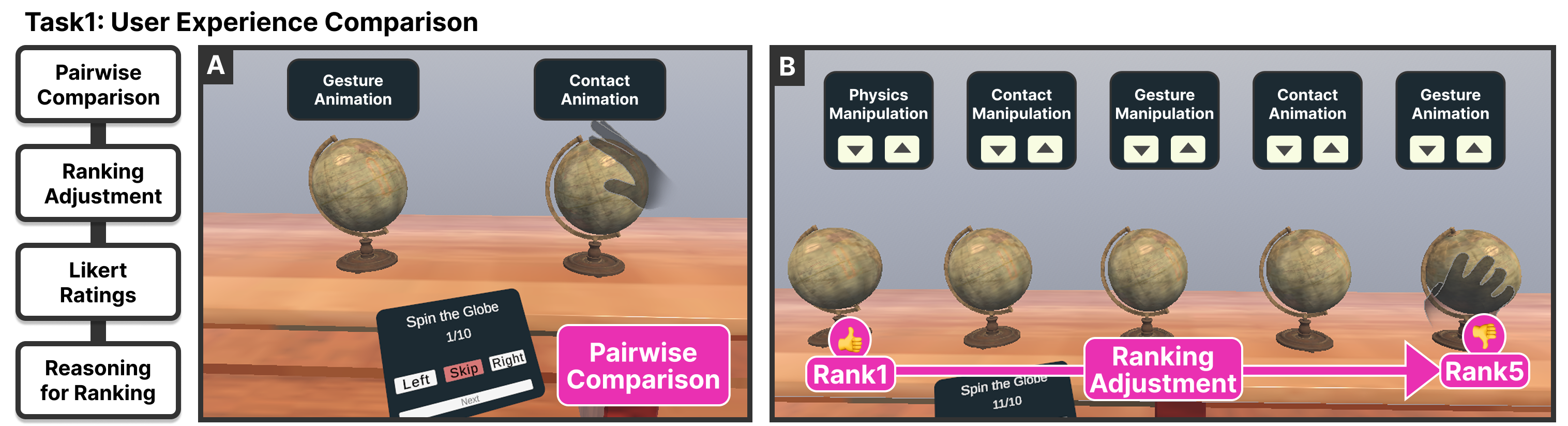}
   \hfil
\caption{Task 1 overview. (A)~Participants compared HOI designs in pairwise trials. (B)~After completing all comparisons, they reviewed and adjusted the ranking. Finally, they rated each HOI on Likert scales and explained their choices through think-aloud.}
\Description{A flowchart and VR screenshots describing Task 1. The flowchart on the left outlines the steps: Pairwise Comparison, Ranking Adjustment, Likert Ratings, and Reasoning. Panel A shows the Pairwise Comparison interface in VR with two globes and buttons for 'Left', 'Skip', and 'Right'. Panel B shows the Ranking Adjustment interface with five globes representing different interaction techniques, allowing users to reorder them using up and down buttons.}
\label{fig:Study1Task1Overview}
\end{figure}

The first task involved pairwise comparisons of HOI designs, as adopted in prior work~\cite{JuicyHapticDesign}, which helps reduce cognitive load compared to evaluating all designs simultaneously. In each stage, we showed participants two versions of the same part with different HOI designs, and asked them to choose their preferred design, where they could skip if the comparison was too difficult (Figure~\ref{fig:Study1Task1Overview}-A). Here, participants could freely manipulate and compare the two versions, and we displayed the applied HOI design above each version to ensure they clearly understood how to perform the interaction.

All possible pairs ($_5C_2 = 10$) were presented in randomized order. Both the sequence of pairs and the within-pair presentation (A/B vs. B/A) were counterbalanced to control for order effects. After completing all pairwise comparisons, the overall HOI design ranking was computed using the round-robin row-sum score procedure~\cite{RoundRobinRanking}. Then, we showed participants the ranking and allowed them to adjust it if desired (Figure~\ref{fig:Study1Task1Overview}-B). We recorded the final confirmed ranking as \textit{Preference} data. Following this, we gathered ratings on a 7-point Likert scale related to key metrics: \textit{Usability} (ease of use: \textit{“This interaction is easy”}, learnability: \textit{“I learned this interaction quickly”}) and \textit{Realism} (realism: \textit{“This interaction is realistic”}). We also collected the reasons behind their preference order. This procedure was repeated for all 13 object-part pairs.

\subsubsection{Task 2: Performance Comparison in Direct Manipulation}

\begin{figure}[!b]
\centering
   \includegraphics[width=\linewidth]{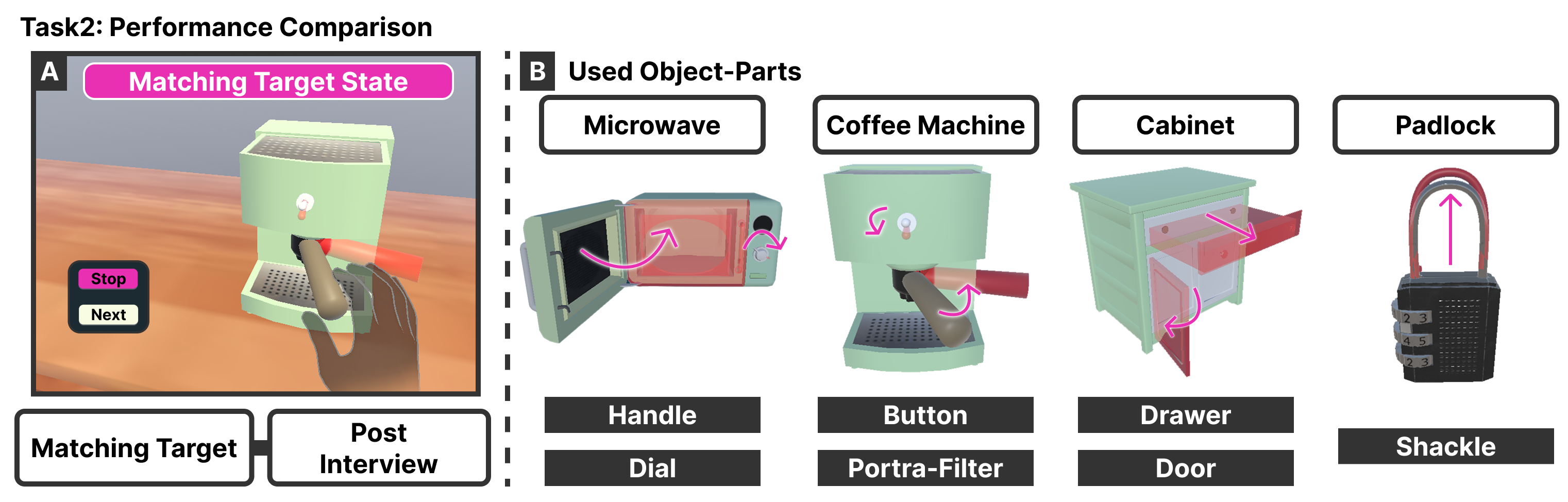}
   \hfil
\caption{Task 2 overview. (A)~Participants matched object-part pairs to target states using different HOI designs~(\textbf{PM, GM, CM}), followed by a post interview. (B) We used a microwave (handle, dial), a coffee machine (button, portafilter), a cabinet (drawer, door), and a padlock (shackle). Red-colored parts indicate the target state to be achieved.}
\Description{A flowchart and VR screenshots describing Task 2. Panel A shows a VR scene with a coffee machine where the user must match a target state highlighted in red, with 'Stop' and 'Next' buttons. Panel B displays the four objects used: a microwave (handle, dial), coffee machine (button, portafilter), cabinet (drawer, door), and padlock (shackle), with red highlights indicating the target state and pink arrows showing motion constraints.}
\label{fig:Task2Overview}
\end{figure}

The second task involved matching the target state, manipulating parts to align them with a red-highlighted target state~(Figure~\ref{fig:Task2Overview}-A). To minimize perceptual errors, the target was overlaid directly on the corresponding part. Participants performed tasks on four multi-part objects covering diverse interaction types~(Figure~\ref{fig:Task2Overview}-B), with certain parts restricted when their motions lacked sufficient salience (e.g., in-place rotation). 

Before starting the task, we showed participants the target state to ensure familiarity so that the comparison across HOI designs was not influenced by target understanding time. Then, we instructed them to complete the task as quickly and accurately as possible. To avoid learning effects, the order of the three HOI designs (\textbf{PM, GM, CM}) was counterbalanced across participants (3! = 6 permutations). Each participant completed two sessions with randomized object order, and performance measures were averaged across sessions to mitigate accidental errors. Each HOI design was tested 8 times per participant (4 objects~$\times$~2 sessions). 

We measured task completion time as the interval between the first and last part manipulation. In addition, we continuously logged the manipulated part’s position at each frame to calculate the reversal count, which is the number of times the error distance changed direction while approaching the target. A lower reversal count indicates greater stability in the HOI design, whereas a higher count suggests repeated re-adjustments. Finally, the last recorded position of manipulated parts was used to compute the error deviation from the target state. These measures were used to compare \textit{Efficiency} across HOI designs. After completing the task, we conducted a semi-structured interview to explore participants’ subjective impressions of \textit{Challenge} in the HOI designs.

\subsection{Results}
\label{sec:Study1Result}

\subsubsection{Data Analysis Procedure}
For both tasks, as the data did not meet the assumptions of normality and homoscedasticity, we applied the Friedman test as a non-parametric method. If the Friedman test was significant, Wilcoxon signed-rank tests were conducted for all method pairs, with the false discovery rate controlled using the Benjamini–Hochberg procedure. 

\subsubsection{Overall Results for Task 1}
Both \textit{Preference} rankings and Likert ratings (\textit{Usability}, \textit{Realism}) revealed clear part-specific variations, indicating the need for HOI mapping to each part rather than applying a uniform design. As shown in Table~\ref{tab:Study1RankingFrideman}, except parts 12 and 13, there is a significant difference between HOI designs across parts ($2-5,8,10-11;p<.001$, $6-7;p<.005, 1,9;p<.05$). Further analysis using the pairwise Wilcoxon signed-rank test revealed that the dominant HOI design varied by part, indicating that no single technique was consistently superior. Similar part-specific variations were also observed in the Likert ratings collected during Task 1, with detailed statistics provided in Appendix~\ref{appendix:Study1Task1Likert}.

\begin{table}[b]
\caption{Friedman test results with Kendall's W and ranking significance. Methods were ordered by their mean score (for Likert scales, higher values indicate better performance; for rank data, lower mean ranks indicate better performance). We then derived a statistically informed grouping: methods not significantly different were grouped together (“=”), while significant differences led to a new, lower-ranked group (“>”). If the Friedman test was not significant, all methods were treated as tied. Significance levels are indicated: * $p<0.05$, ** $p<0.01$, *** $p<0.005$, **** $p<0.001$).}
\begin{tabular}{c c l l}
\multicolumn{4}{l}{\small $n=20,\ k=5$} \\
\hline
Part & Kendall's W & Friedman\_sig & Ranking\_sig \\
\hline
1 & 0.1290 & p=0.035*    & \textbf{CA=CM=GA}>GM=PM \\
2 & 0.2365 & p<0.001**** & \textbf{CM=CA}>GM=GA=PM \\
3 & 0.4010 & p<0.001**** & \textbf{CM=GM}>PM=CA=GA \\
4 & 0.2490 & p<0.001**** & \textbf{CM=CA}>GM=GA=PM \\
5 & 0.3950 & p<0.001**** & \textbf{CM}>GM=CA>GA=PM \\
6 & 0.1980 & p=0.003***  & \textbf{CM=GM=CA}>GA=PM \\
7 & 0.1955 & p=0.003***  & \textbf{CM=CA=PM}>GM=GA \\
8 & 0.2295 & p=0.001*** & \textbf{CA=GM=GA=CM}>PM \\
9 & 0.1575 & p=0.013*    & \textbf{CM=GM=PM=CA}>GA \\
10 & 0.5365 & p<0.001**** & \textbf{CM=PM}>GM=CA>GA \\
11 & 0.3100 & p<0.001**** & \textbf{CM=CA}>GA=PM=GM \\
12 & 0.0980 & p=0.097   & \textbf{PM=CM=GM=CA=GA} \\
13 & 0.0265 & p=0.713   & \textbf{CM=CA=GM=GA=PM} \\
\hline
\end{tabular}
\label{tab:Study1RankingFrideman}
\Description{Results of the Friedman test for Study 1 Task 1 rankings. The table reports Kendall’s W for agreement strength and identifies statistically significant differences in HOI design rankings.}
\end{table}

\label{sec:study1task1qualitative}
Based on participants’ think-aloud explanations, their preferences were strongly shaped by real-world habits of object manipulation. Gesture-based methods were favored when the required gesture aligned with everyday use (e.g., grasping a microwave handle to open it), but were seen as inconvenient or unrealistic when they conflicted with natural habits (e.g., rotating a globe with a prescribed gesture rather than simply spinning it by hand). Importantly, these habitual patterns were not arbitrary but closely reflected object-part characteristics such as affordances, size, and motion constraints. Furthermore, participants preferred direct manipulation for fine-grained or continuous control (e.g., scissors, sprayer, dials), while animation was valued for comfort and clear state transitions in large or discrete motions (e.g., doors, laptop lids). These findings suggest that preferences for HOI designs are fundamentally grounded in part characteristics, which enables us to generalize to new parts by aligning them with similar ones in our dataset (Sec~\ref{sec:HOIMapper}).

\subsubsection{Overall Results for Task 2}
\label{sec:study1task2result}
\begin{figure*}[t]
\centering
   \includegraphics[width=0.8\textwidth]{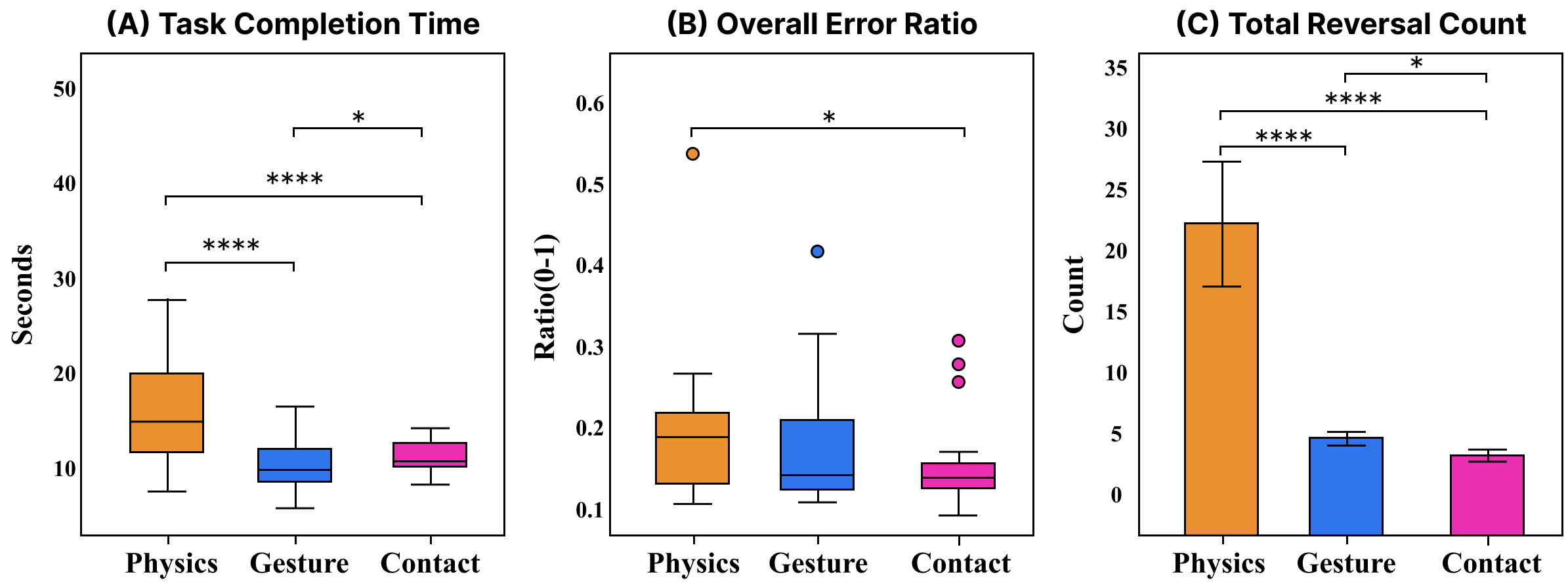}
   \hfil
\caption{Task performance comparison across \fcolorbox{orange}{orange!30}{Physics-(PM)}, \fcolorbox{baselineBlue}{baselineBlue!30}{Gesture-(GM)}, and \fcolorbox{oursPink}{oursPink!30}{Contact-(CM)}based manipulation. The results include (A) task completion time, (B) overall error ratio, and (C) total reversal counts: * $p<0.05$, ** $p<0.01$, *** $p<0.005$, **** $p<0.001$.}
\Description{Three boxplots comparing performance metrics across Physics (PM), Gesture (GM), and Contact (CM) manipulations. Panel A (Task Completion Time) shows Physics is significantly slower than Gesture and Contact (p < 0.001). Panel B (Error Ratio) shows Physics generally has a higher error ratio than Contact (p < 0.05). Panel C (Reversal Counts) indicates Physics has significantly higher reversal counts compared to both Gesture and Contact (p < 0.001).}
\label{fig:Study1Task2}
\end{figure*}
Figure~\ref{fig:Study1Task2} illustrates the overall results for Task 2. We compared task completion time, error ratio, and reversal count as key indicators of \textit{Efficiency} across HOI designs. Overall, the results showed that \textbf{PM} consistently underperformed compared to the others, suggesting that in terms of \textit{Efficiency} the practical choice lies between \textbf{GM} and \textbf{CM}. In addition, post-interview findings revealed that \textit{Challenge} was primarily differentiated between \textbf{GM}, which required mastery of gestures, and \textbf{PM}, which demanded fine control under unpredictable responses.

 A Friedman test for task completion time revealed significant differences across conditions ($p<.001$). Follow-up Wilcoxon signed-rank tests showed that \textbf{PM} was slower than both \textbf{GM} and \textbf{CM} ($p<.001$), and that \textbf{CM} was faster than \textbf{GM} ($p<.05$), as shown in Figure~\ref{fig:Study1Task2}-A. Error ratio was measured by normalizing final positional and rotational deviations and combining them with equal weights. Figure~\ref{fig:Study1Task2}-B shows that no significant differences were observed across HOI designs (overall error ratio: $p=.157$). This suggests that participants were willing to trade speed for accuracy, achieving comparable precision across conditions even when task completion times differed. In particular, \textbf{PM} exhibited greater variability, as fine control was often difficult. For example, participants sometimes resorted to “slamming” the microwave door shut rather than aligning it precisely. As shown in Figure~\ref{fig:Study1Task2}-C, reversal counts differed significantly across HOI designs ($p<.001$). Pairwise comparisons showed that \textbf{PM} produced significantly more reversals than both the \textbf{GM} and \textbf{CM} ($p<.001$), and that \textbf{GM} produced more reversals than \textbf{CM} ($p<.05$). This latter difference was most notable for parts lacking clear grasping affordances, such as the drawer’s bottom door without a distinct handle.

Building on the quantitative findings where \textbf{PM} already showed lower \textit{Efficiency}, post interviews highlighted more nuanced trade-offs between \textbf{GM} and \textbf{CM}. \textbf{GM} was valued for precise intent alignment~(P1,P10–11,P13–14,P16,P19), though some noted physical fatigue~(P6–7,P19). \textbf{CM} was consistently described as easy and comfortable~(P1–2,P9,P12), yet prone to unintentional manipulation~(P15–16,P18–19). Overall, these findings highlight a trade-off: \textbf{GM} offers precision and control, whereas \textbf{CM} provides ease and comfort, suggesting that the appropriate choice depends on the primary design focus.

Participants also described different forms of \textit{Challenge}, primarily contrasting \textbf{GM} and \textbf{PM}. For \textbf{GM}, the challenge stemmed from developing mastery, as participants felt they needed practice to reliably perform gestures: \textit{``I had to get used to which gesture to use and how to do it well''~(P9)}, \textit{``The more I refined the gesture, the better the outcome seemed''~(P17)}. For \textbf{PM}, the difficulty lay in fine physical control and unpredictable responses: \textit{“Fine control was hard and I had to carefully adjust the force”~(P1, P6, P8)}, \textit{``The outcome was unexpected but when it finally matched, it felt rewarding''~(P14, P20)}. These responses suggest that \textbf{GM} elicited a skill-based, mastery-oriented challenge, while \textbf{PM} posed a control-based, execution-oriented challenge. By contrast, \textbf{CM} was rarely described as challenging, reinforcing its role as the most comfortable method.

\subsection{System Design Implications}
\label{sec:study1findings}
Task~1 showed that \textit{Preference}, \textit{Usability}, and \textit{Realism} varied across object parts, motivating part-level preference mapping. Task~2 further highlighted binary trade-offs: \textbf{GM vs.\ CM} for \textit{Efficiency}, and \textbf{GM vs.\ PM} for \textit{Challenge}. We formulate our HOI mapping module based on these findings, which supports both \textit{ranking-based} and \textit{binary-decision} modes (Sec.~\ref{sec:HOIMapper}).

Our study further revealed opportunities to refine HOI designs through adjustable parameters. We initially set default parameter values and refined them through a pilot study to ensure baseline usability. Nonetheless, participants reported variations in comfort and control, indicating individual differences in how these defaults were experienced. Their feedback often pointed to specific limitations, such as unintended movements in physics-based manipulation, premature deselection in gesture-based methods, or unnatural step sizes in animation. Although expressed as user frustrations rather than explicit design requests, these observations highlighted the need for incorporating customization capability for the user. Thus, we added a customization module that enables designers to configure factors such as resistance, gesture set, release distance, animation mode, and step angle, as detailed in Section~\ref{sec:HOIDesignCustomization}.

\section{\system}
\begin{figure*}[t]
\centering
   \includegraphics[width=\textwidth]{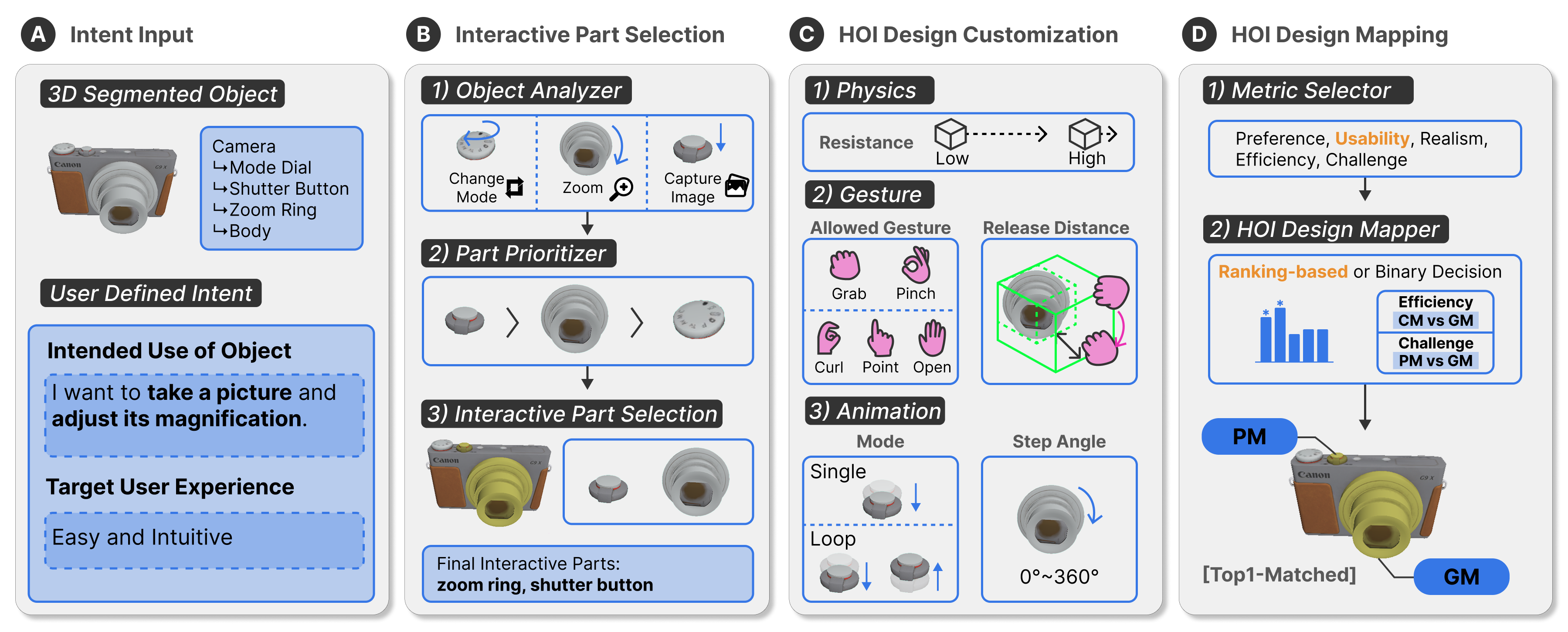}
   \hfil
\caption{\system{} system workflow. Given a 3D object and a design intent, the system first analyzes the 3D object to recommend interactive parts. Then, designers customize HOI properties (e.g., physics, gestures, and animation) and finalize the mapping through ranked recommendations.}
\Description{A system pipeline diagram with four components. 1) Intent Input takes a 3D object and designer intent. 2) Interactive Part Selection uses an object analyzer and priority selector to identify parts. 3) HOI Design Customization allows tuning of Physics, Gesture, and Animation properties. 4) HOI Design Mapping selects metrics like usability and recommends the final HOI design using ranking or binary decisions.}
\label{fig:SystemOverview}
\end{figure*}

We developed \system{}, a VLM-based in-situ authoring tool that supports designers to create part-level interactive objects in VR, guided by their intended interaction goals. \system{} consists of three main modules~(see Figure~\ref{fig:SystemOverview}). These modules include (1) interactive part selection, (2) HOI design customization, and (3) HOI design mapping.

We offered these features in an in-situ environment to support rapid prototyping with refinement capability. The authoring workflow begins with the user providing a target 3D object (which is segmented into parts, labeled with part names, and pre-defined with motion constraints) and specifying a design intent as text (the intended use of the object and target user experience). 

The hardware configurations and VR deployment used in \system{} were consistent with those detailed in Section~\ref{sec:Implementation}. The system incorporates a multimodal LLM backend (GPT-4o~\cite{achiam2023gpt}), leveraging its built-in vision–language capabilities for object analysis. The detailed system prompts used in \system{} are provided in Appendix~\ref{appendix:SystemPrompts}.

\subsection{Interactive Part Selection}
Before defining HOI designs, designers must first decide which parts of an object should be interactive. \system{} supports this process by analyzing object characteristics and the intended use, generating prioritized recommendations of interactive parts, and allowing designers to override them.

\subsubsection{Object Analyzer}
This module combines visual inputs and part hierarchies, using a VLM to infer interaction types (e.g., rotate, slide) and functional affordances (e.g., take a picture, change magnification) for each part. Here, visual input is obtained by normalizing the object’s scale to a reference cube and capturing multi-view images with 8 cameras (4 above and 4 below) placed at 90-degree intervals around the object center, ensuring full coverage and consistent scale. These raw inputs are then transformed into a structured representation, \textbf{[object–part–interaction–affordance]}, which serves as the basis for the following modules. Since VLM inference is costly, it is performed only once at the start, and the representation is leveraged by subsequent LLM modules to support real-time reasoning throughout the authoring pipeline.

\subsubsection{Part Prioritizer}
This module analyzes the given intended use of the object together with part information obtained from \textit{Object Analyzer} and produces a prioritized list of all parts, ranking them by their relevance to the intended use. As the number of parts increases, manually selecting them becomes increasingly difficult. Therefore, the system leverages the LLM to recommend an initial number of interactive parts, providing designers with a suitable starting point. The prioritized list and recommended number of interactive parts are then passed to the subsequent Interactive Part Selection process.

\subsubsection{Interactive Part Selection}
Designers can then refine the recommended interactive parts in two ways: (1) by specifying a desired number of interactive parts, in which case the system automatically activates the corresponding top-ranked parts from the list, or (2) by manually selecting parts to activate when the priority does not fully match their intent. This dual approach balances between design automation and flexibility. 

\subsection{HOI Design Customization}
\label{sec:HOIDesignCustomization}
Drawing on the findings from Section~\ref{sec:study1findings}, we define key parameters that allow designers to customize HOI designs. The resulting customized configuration is then integrated into the final HOI Mapping process described in the next section.

\subsubsection{Physics}
Designers can adjust \textbf{Resistance}, linked to the rigid body’s resistance value. Here, higher values mean more force to continue movement, and lower values produce a more responsive feel~(Figure~\ref{fig:SystemOverview}-C-1).

\subsubsection{Gesture}
For gesture criteria, designers can adjust two key parameters~(Figure~\ref{fig:SystemOverview}-C-2): 

\begin{enumerate}
    \item \textbf{Allowed Gesture Set:} Designer can choose the type of gesture to acquire and maintain control (Grab, Pinch, Curl, Point, and Open). We referred to a gesture set from prior taxonomies~\cite{karam2005taxonomy}, design guidelines~\cite{xia2022iteratively}, and Meta’s default API%
\footnote{\url{https://developers.meta.com/horizon/documentation/unity/unity-isdk-detecting-poses}}$^{,}$\footnote{\url{https://developers.meta.com/horizon/documentation/unity/unity-isdk-grabbing-objects}}.
 These gestures serve as selection triggers. Compound gestures that occur only after acquisition (e.g., squeeze and swipe) were excluded since they address post-selection actions rather than the selection itself.
    \item \textbf{Release Distance:} Designer can define when control is released as the hand moves away (small values for quick release, large values for sustained control). Release distance is defined relative to the object’s mesh boundary. At acquisition, the system records the part’s trigger center as an anchor. The distance between this anchor and the user’s index fingertip is continuously measured and normalized to the part’s scale. When this distance exceeds the threshold set by the designer, control is released.
\end{enumerate}

\subsubsection{Animation}
For animating given 3D objects, designers can configure two parameters~(Figure~\ref{fig:SystemOverview}-C-3): 
\begin{enumerate}
    \item \textbf{Animation mode:} For constrained parts (e.g., doors with open–close limits), designers can choose between Single mode, which allows the part to move in one-way direction to the motion range limit (e.g., from fully closed to fully open), and Loop mode, which allows the part to move back and forth. 
    \item \textbf{Step angle:} For unconstrained parts (e.g., a 360° dial), interactions progress in increments of the specified angle, supporting both coarse and fine adjustments.
\end{enumerate}

\subsection{HOI Design Mapping}
\begin{figure}[b]
\centering
   \includegraphics[width=\linewidth]{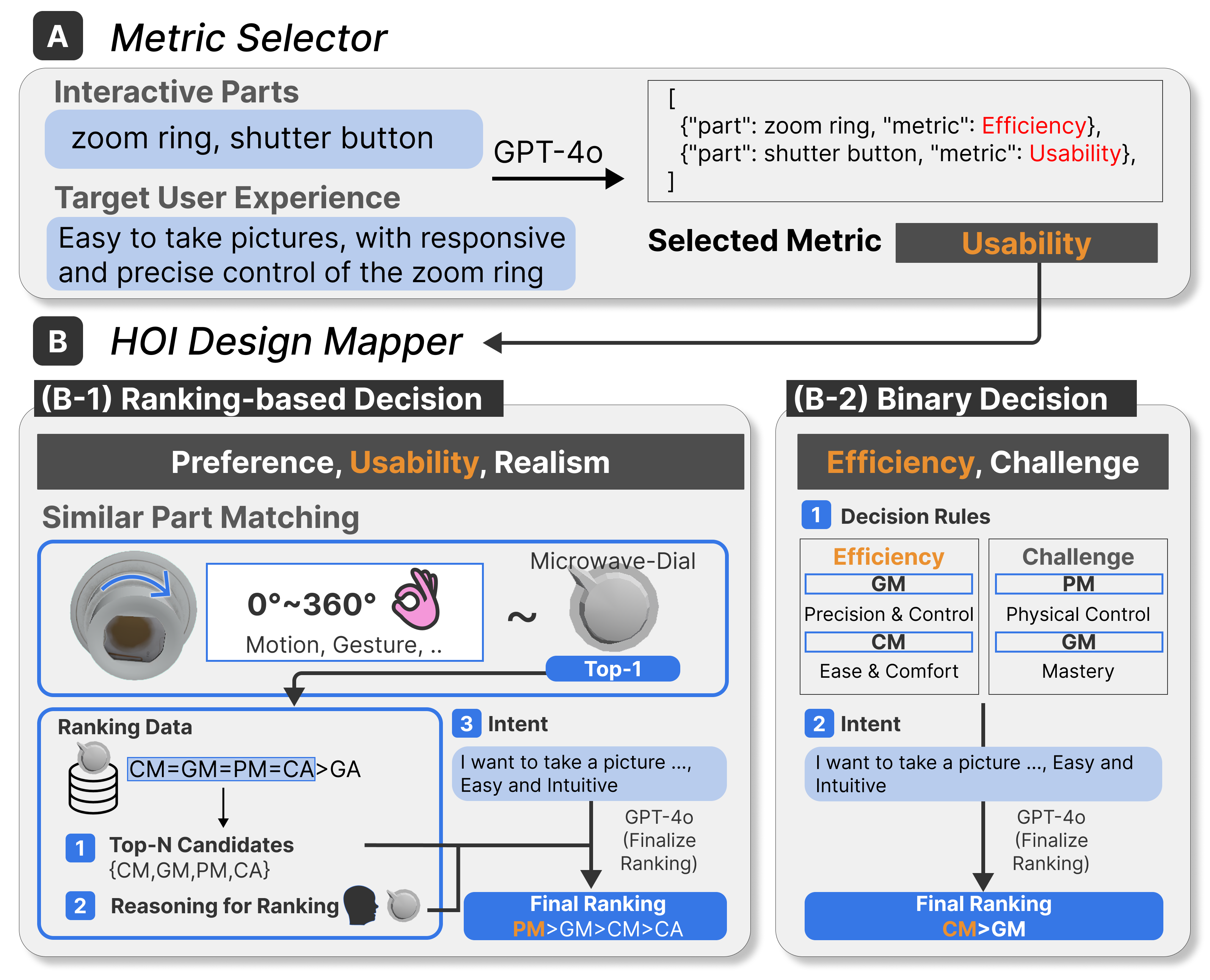}
   \hfil
\caption{The detailed workflow of the HOI mapping module in \system{}. (A)~Metric Selector interprets the target user experience and assigns it to one of five metrics. (B)~HOI Design Mapper determines a ranked list for HOI designs based on selected metric, using (B-1)~ranking-based decision for \textit{Preference}, \textit{Usability}, and \textit{Realism} and (B-2)~binary decision for \textit{Efficiency} and \textit{Challenge}. The top-1 result is applied by default, while the top-$N$ list supports further exploration.}
\Description{A detailed workflow diagram of the HOI mapping module. The Metric Selector maps user experience goals to metrics like usability. The HOI Design Mapper then determines the design: for preference/usability/realism, it uses a ranking-based decision with GPT-4o; for efficiency/challenge, it uses a binary decision based on predefined rules. The output is a finalized ranking of HOI designs.}
\label{fig:HOIMapping}
\end{figure}

Figure~\ref{fig:HOIMapping} illustrates the detailed process of the module. This module determines the most appropriate HOI design for the given intent. To this end, we first link the intent to a key metric through the \textit{Metric Selector}, and then apply the corresponding decision mode in the \textit{HOI Design Mapper} to produce the final mapping.

\label{sec:hoimapping}

\subsubsection{Metric Selector~(Figure~\ref{fig:HOIMapping}-A)}
\label{sec:MetricSelector}
This module interprets the designer’s intent and maps it to one of four metrics~(\textit{Realism, Usability, Efficiency, and Challenge}), defaulting to \textit{Preference} when no clear match is found. To prevent overfitting, we adopted a strict keyword-based framework: intents are mapped to a metric only when they explicitly include associated terms (e.g., “realistic physics” → realism, “easy for beginners” → usability). The complete keyword list is provided in Appendix~\ref{appendix:metric_selector}. Efficiency was further anchored to task-related terms such as speed, accuracy, or error rate, consistent with our performance measures (Sec.~\ref{sec:study1task2result}). This ensures that \textit{Preference} remains the default when ambiguous signal is present.

\subsubsection{HOI Design Mapper~(Figure~\ref{fig:HOIMapping}-B)}
\label{sec:HOIMapper}
This module integrates the findings from Section~\ref{sec:study1findings}. It applies ranking-based decisions for \textit{Preference, Realism} and \textit{Usability}, and binary decisions for \textit{Efficiency} and \textit{Challenge}. The choice of decision mode is determined by the selected metric from \textit{Metric Selector}. 

\begin{itemize}
\item \textbf{Ranking-based Decision~(Figure~\ref{fig:HOIMapping}-B-1).} This mode begins by matching the given part to the most similar part in the dataset. Since our dataset was constructed to capture distinctive part characteristics, the system selects a single top match. The matching criteria include gesture type, physical scale, motion constraints, and control granularity, all of which shape user preferences at the part level (Sec~\ref{sec:study1task1qualitative}). To capture these nuances, the system employs an LLM-based approach rather than simple text similarity. Once the system identifies the best matching part, it retrieves ranking data for selected metric and draws initial candidates from the top-ranked tiers (e.g., $\mathcal{C}_{\text{init}} = \{CA, CM, GA\}; CA=CM=GA>GM=PM$, Sec~\ref{sec:Study1Result}), ensuring that clearly favored groups are prioritized. The LLM then decides the order within these tiers by leveraging ranking rationales from Section~\ref{sec:5DataCollection}. Using this contextual input, the system produces a prioritized ordering that aligns empirical findings with the designer’s intent. The output consists of both a top-1 suggestion for direct mapping and a top-$N$ list for flexible design exploration.
\item \textbf{Binary Decision~(Figure~\ref{fig:HOIMapping}-B-2)}. In this mode, binary HOI design choices are encoded into the prompt with decision rules derived from Section~\ref{sec:study1findings}. \textbf{GM} vs.\ \textbf{CM} for \textit{Efficiency} and \textbf{GM} vs.\ \textbf{PM} for \textit{Challenge} are represented as binary options, and the LLM selects the option that best aligns with the designer’s intent.
\end{itemize}

\subsection{Authoring Interface}
\begin{figure*}[t]
\centering
   \includegraphics[width=0.9\textwidth]{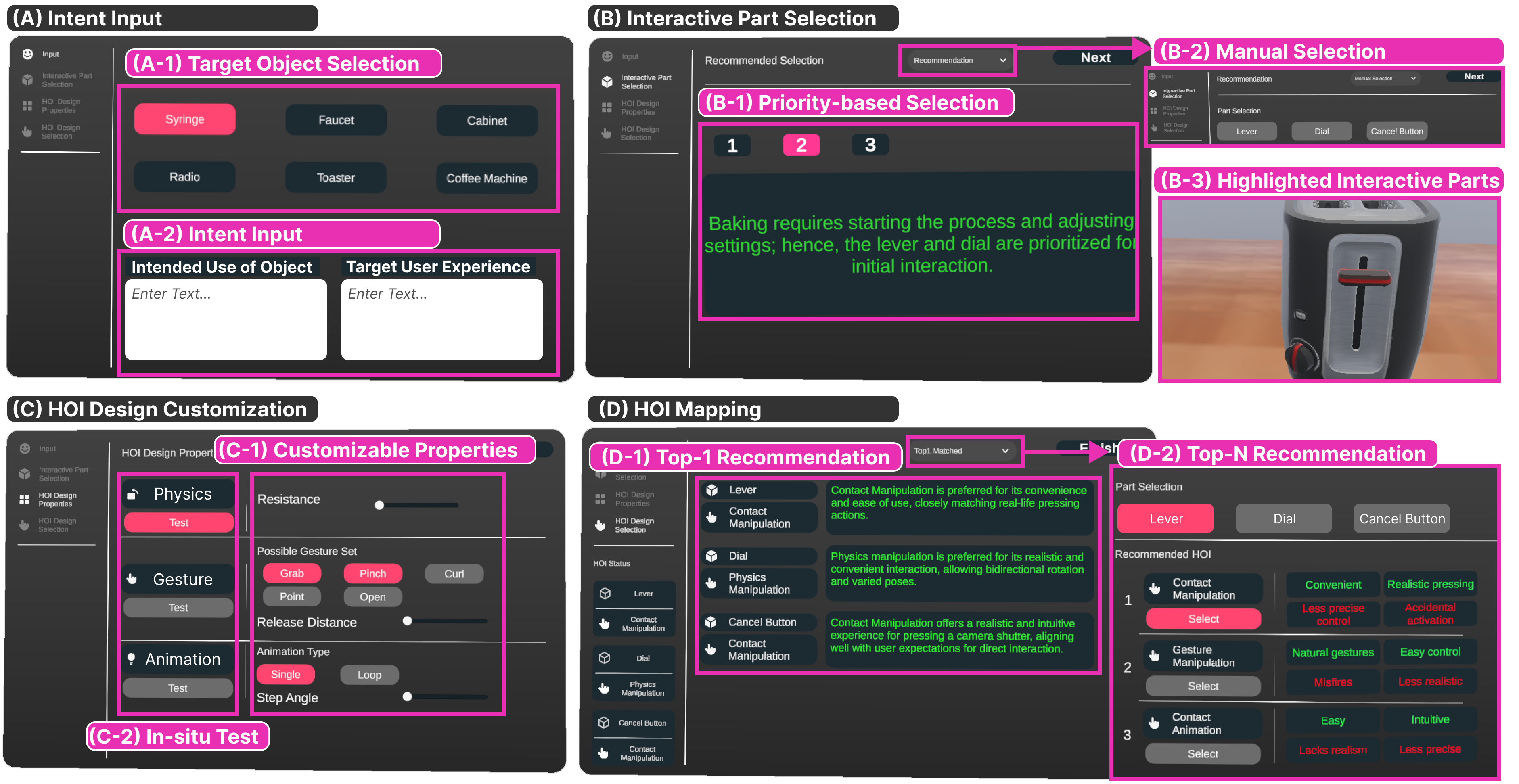}
   \hfil
\caption{Interface overview of \system. (A)~\textbf{Intent Input}. The system loads the pre-processed object (segmented, labeled, and motion defined), and designers select the target object using a toggle~(A-1). Then, designers specify the intended use of the object and the target user experience with text input~(A-2). (B)~\textbf{Interactive Part Selection}. The system selects interactive parts based on intent, which can be refined through two modes: (B-1)~priority-based selection or~(B-2)~manual selection. Selected parts are visually emphasized with red highlights~(B-3). (C)~\textbf{HOI Design Customization}. Designers adjust~(C-1) properties and validate changes in-situ ~(C-2). (D)~\textbf{HOI Mapping}. The system first presents a top-1 recommendation~(D-1), while additional top-$N$ alternatives with pros and cons are provided for further refinement~(D-2).}
\Description{A screenshot of the authoring interface showing four panels. Panel A (Intent Input) allows object selection and text input for design intent. Panel B (Interactive Part Selection) offers priority-based or manual selection of parts, with highlights on the 3D model. Panel C (HOI Design Customization) provides sliders for physics, gesture, and animation settings. Panel D (HOI Mapping) displays the top recommendation with a rationale and a dropdown for alternative options.}
\label{fig:AuthoringInterface}
\end{figure*}

We designed a VR authoring interface that integrates the functionalities described in the previous sections (Figure~\ref{fig:AuthoringInterface}). The interface guides designers through four main steps, enabling in-situ authoring that balances automation with manual control. Throughout the workflow, a side panel indicates the current step, and designers can proceed by clicking the next button at the top.

\begin{enumerate}
    \item \textbf{Input Intent~(Figure~\ref{fig:AuthoringInterface}-A)}, which allows designers to specify their design intent (intended use of the object and target user experience) by typing with a virtual keyboard. 
    \item \textbf{Interactive Part Selection~(Figure~\ref{fig:AuthoringInterface}-B)}, which highlights selected interactive parts~(Figure~\ref{fig:AuthoringInterface}-B-3). Designers can refine these either through (1) priority-based selection~(Figure~\ref{fig:AuthoringInterface}-B-1), where selecting a number automatically activates the corresponding top-ranked parts, or (2) manual selection~(Figure~\ref{fig:AuthoringInterface}-B-2), where parts are directly toggled on or off. A dropdown menu allows switching between the two modes.
    \item \textbf{HOI Design Customization~(Figure~\ref{fig:AuthoringInterface}-C)}, which provides adjustable parameters (resistance, gesture set, release distance, animation mode, and step angles) with sliders and toggles~(Figure~\ref{fig:AuthoringInterface}-C-1). Each design can be immediately tested in-situ~(Figure~\ref{fig:AuthoringInterface}-C-2); for example, release distance is visualized as a surrounding sphere, and other parameters can be directly manipulated within the VR scene.
    \item \textbf{HOI Design Mapping~(Figure~\ref{fig:AuthoringInterface}-D)}, which presents a top-1 recommendation with a brief rationale~(Figure~\ref{fig:AuthoringInterface}-D-1). Designers can also explore top-$N$ alternatives via a dropdown menu and refine their choice based on the provided pros and cons~(Figure~\ref{fig:AuthoringInterface}-D-2).
\end{enumerate}

\section{Study 2: Evaluating \system{} with HOI Authoring} 
\label{sec:userstud}

To evaluate the effectiveness of \system{}, we conducted a user study with two main tasks. Task 1 focused on the HOI design recommendation module in isolation, comparing it against a manual baseline to assess efficiency and exploration behavior. Task 2 evaluated the full HOI authoring workflow through a usability test and how all core modules work together in practice.

\subsection{Baseline Design}
\begin{figure}[t]
\centering
  \includegraphics[width=\linewidth]{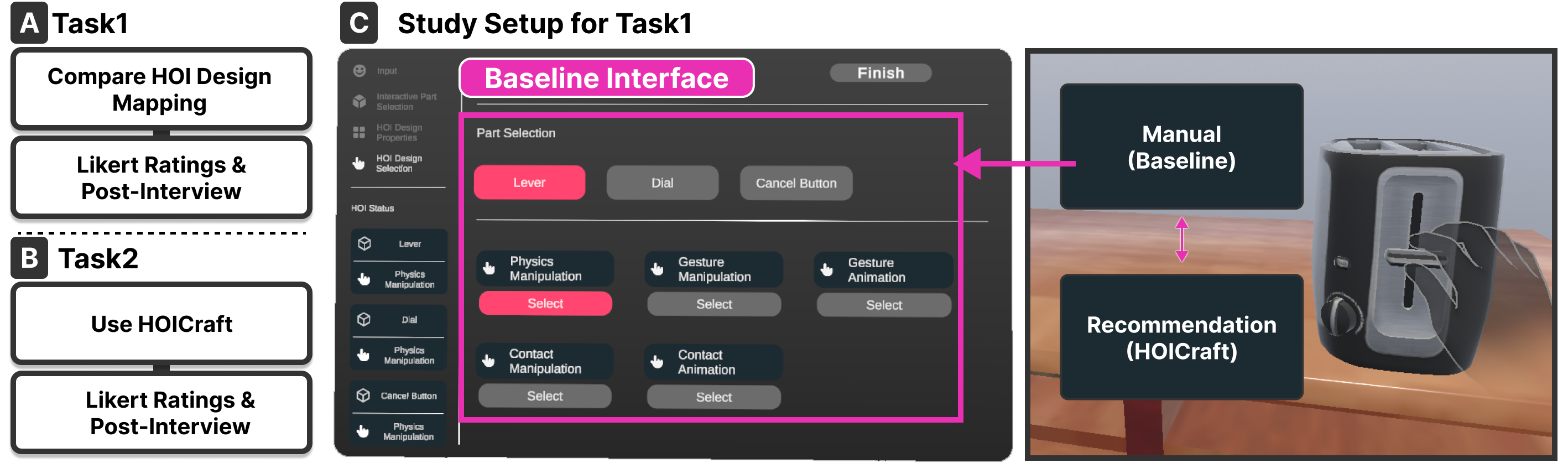}
   \hfil
\caption{Study 2 overview. (A) Task 1: Participants designed HOIs for six objects to compare the baseline manual mapping interface~(C) with our recommendation-based mapping interface~(Figure~\ref{fig:AuthoringInterface}). The different object sets (3 objects) were used for baseline and recommendation conditions with Likert-scale rating and post-interview. (B)~Task 2: Participants used our system to author HOIs for 4 objects through the full workflow, followed by a post-task interview.}
\Description{An overview of User Study 2. The left diagram shows the procedure for Task 1 (Comparison) and Task 2 (Full Workflow). Panel C shows the Task 1 setup comparing a 'Baseline Interface' (manual selection buttons) against the 'Ours' approach (recommendation-based), illustrated with a toaster object.}
\label{fig:Study2Overview}
\end{figure}

Figure~\ref{fig:Study2Overview}-C illustrates the baseline interface. We designed the baseline condition to reflect how designers would work without automated support. Here, users tested HOI designs one by one through trial-and-error and manually choose which to adopt. In contrast, our system condition provided prioritized recommendations and rationales based on user preference data. This setup allowed us to investigate not only whether AI assistance could reduce cognitive and physical burdens in mapping HOI designs, but also whether the outcomes were perceived as comparable in quality to manually created ones.

\subsection{Study Setup}
We recruited 12 VR interaction designers in the research field (5 female), aged 23 to 30 ($M=26.08$, $SD=2.31$). Based on a demographics survey, they had VR interaction design/development experience ($M= 2.04$, $SD=1.17$) and Unity development experience ($M= 2.08$, $SD=0.99$). Each study session lasted about 1.5 hours. The study was approved under the IRB, and participants received compensation for their time with 15 USD. The hardware configurations and VR deployment used in the study were consistent with those detailed in Section~\ref{sec:Implementation}. The study consisted of three phases: (1) \textit{Introduction and Training}, (2) \textit{Task Session}, and (3) \textit{Post Interview}.

\subsubsection{Introduction and Training}
We first collected participants' demographic information and explained the study tasks. Next, we introduced each HOI design, and participants were encouraged to ask questions about them freely, as we assumed they were already familiar with them. We informed their role was to design HOIs aligned with specific target user experiences~(easy \& intuitive, realistic, and general preference). We also explained the two interface conditions: the manual selection (baseline) and recommendation-based selection (our system). After the explanation, participants completed a training session with both interfaces. During this session, they practiced each HOI design to gain an understanding of its characteristics. Training continued until participants reported that they could clearly distinguish among HOI designs and understood how to operate the interfaces, usually within five minutes.

\subsubsection{Task 1: HOI Mapping Comparison}
Participants compared the baseline (manual mapping) with our recommendation-based mapping by creating HOIs for 6 objects (3 in the manual condition and 3 in the recommendation condition). Each object set contained different levels of complexity based on the number of interactive parts, including simple (e.g., faucet and syringe), medium (e.g., cabinet and radio), and complex (e.g., toaster and coffee machine) objects. In each trial, a target user experience was randomly assigned, and participants created HOIs aligned with it.

To prevent learning or carryover effects, we prepared two distinct sets of objects (Set A and Set B). Group 1 completed Set A with the manual interface and Set B with the recommendation interface, while Group 2 followed the opposite assignment. Here, the set order itself was fixed (A before B), as our analyses focused on comparing interface versions rather than differences across individual objects. Because both sets included a balanced range of object complexities and were counterbalanced across conditions, each interface was equally likely to be tested in the first or second half of the session. Within each set, the order of objects was randomized to mitigate local sequence effects. We collected two quantitative measures, including decision speed (time from mapping initiation to confirmation) and exploratory count (number of times participants revised or tested different HOI designs). After completing Task 1, participants filled out a questionnaire (Table~\ref{tab:Study2Task1Likert}).

\begin{table}[t]
    \centering
    \caption{Questionnaires for comparative evaluation between baseline and \system{}~(Task 1)}
    \label{tab:Study2Task1Likert}
    \begin{tabular}{p{0.05\linewidth}p{0.75\linewidth}p{0.1\linewidth}}
        \toprule
        \textbf{Item} & \textbf{Questionnaire} & \textbf{Answer Type} \\
        \midrule
        1 & (Ease of Decision) I felt the HOI mapping process was intuitive and easy. & Likert (1-7) \\
        2 & (Decision Efficiency) I felt the HOI mapping process was efficient. & Likert (1-7) \\
        3 & (Decision Confidence) I am confident in the HOI mapping decision I made.  & Likert (1-7) \\
        4 & (Variety of Considerations) I was able to sufficiently consider multiple factors appropriate for the intent. & Likert (1-7) \\
        5 & (Outcome Satisfaction)  I believe the HOI mapping decision I made best aligns with the intended user experience. & Likert (1-7) \\
        \bottomrule
    \end{tabular}
    \Description{Questionnaire items for Study2 Task 1 evaluation, focusing on ease of decision, efficiency, confidence, considerations, and satisfaction.}
\end{table}

\subsubsection{Task 2: Usability Test}
Participants evaluated the overall usability of the \system{} by completing the full HOI authoring workflow: (1) selecting an object and specifying the intended use and target user experience, (2) selecting object parts to be interactive, (3) customizing HOI design properties, and (4) finalizing the HOI mapping. Unlike Task 1, this task focused solely on our system rather than comparing it with the manual baseline, aiming to assess how well it supports usability and reduces workload. To this end, participants completed a Likert-scale survey in Table~\ref{tab:Study2Task2AuthoringEval} related to usability and satisfaction as well as the System Usability Scale (SUS).

\begin{table}[ht]
    \centering
    \caption{Questionnaires for HOI authoring evaluation~(Task 2)}
    \label{tab:Study2Task2AuthoringEval}
    \begin{tabular}{p{0.05\linewidth}p{0.75\linewidth}p{0.1\linewidth}}
        \toprule
        \textbf{Item} & \textbf{Questionnaire Item} & \textbf{Answer Type} \\
        \midrule
        1 & (Interactive Part Selection) The priority list provided by the system was helpful in deciding which part to be interactive. & Likert (1-7) \\
        2 & (HOI Design Mapping) The rationale provided by the system was useful in making a decision. & Likert (1-7) \\
        3 & (HOI Design Mapping) The recommendations aligned well with my intended user experience. & Likert (1-7) \\
        4 & (Authoring) The tool helped me make effective design decisions for diverse intents. & Likert (1-7) \\
        5 & (HOI Design Customization) The customization is effective to fine-tune HOI design to match my intent. & Likert (1-7) \\
        6 & (Outcome Satisfaction) I am satisfied with the quality of the outcome that I created. & Likert (1-7) \\
        7 & (Workflow Satisfaction) I am satisfied with the overall VR HOI design workflow. & Likert (1-7) \\
        \bottomrule
    \end{tabular}
    \Description{Questionnaire items for Study2 Task 2 evaluation, covering part selection, design mapping, customization, authoring, and satisfaction.}
\end{table}

\subsection{Results}

\begin{figure*}[t]
\centering
   \includegraphics[width=\textwidth]{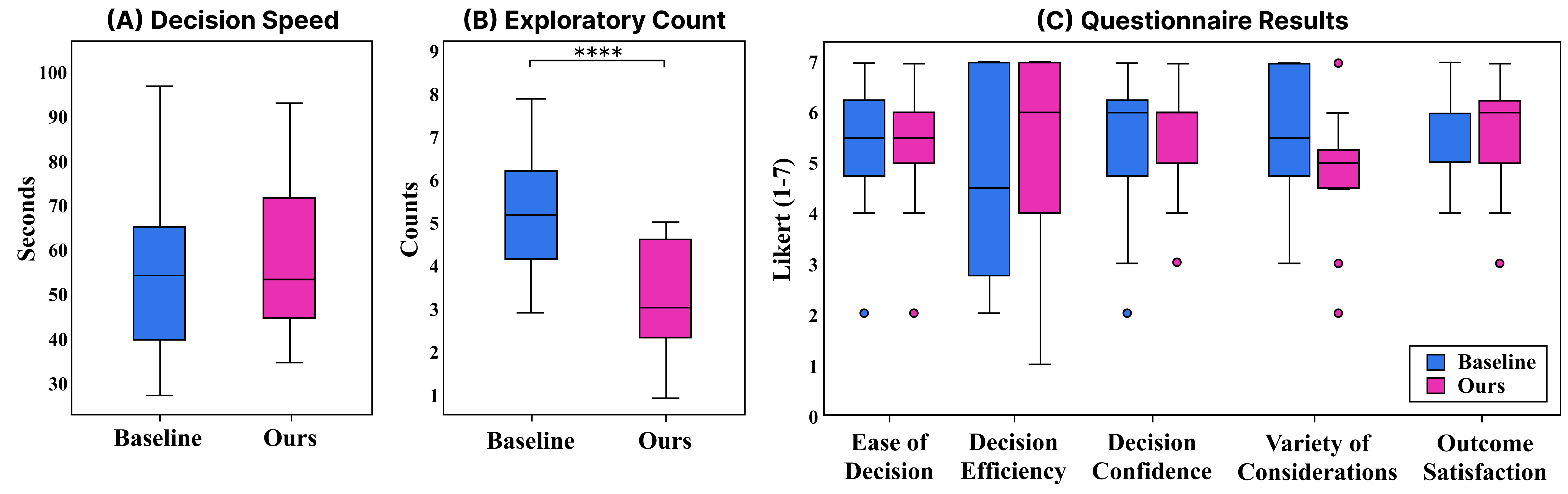}
   \hfil
\caption{Comparison of (a) decision speed, (b) exploratory count, and (c) Likert questionnaire across \fcolorbox{baselineBlue}{baselineBlue!30}{Baseline} and \fcolorbox{oursPink}{oursPink!30}{Ours}: **** $p<0.001$.}
\Description{Boxplots comparing the Baseline and Ours conditions. Panel A shows no significant difference in Decision Speed. Panel B shows significantly fewer Exploratory Counts for Ours compared to Baseline (p < 0.001), indicating reduced trial-and-error. Panel C shows Likert ratings for five subjective metrics, with both conditions achieving comparably high scores.}
\label{fig:Study2Task1Result}
\end{figure*}

\subsubsection{Overall Results for Task 1}
A Wilcoxon signed-rank test showed no significant difference in decision speed between versions (baseline and ours, $p=0.519$) as shown in Figure~\ref{fig:Study2Task1Result}-A, indicating that neither interface offered an overall speed advantage. This reflects a trade-off: the manual interface involved more initial trial-and-error, but participants quickly formed internal rules to speed decisions (P1, P3, P9). In contrast, the recommendation-based interface reduced options but required extra time to read and evaluate rationales. Although speed did not differ in this controlled setting, participants agreed the recommendation-based interface would provide clear advantages at scale. They emphasized that with more design elements, the recommendation-based interface's ability to prune irrelevant options would be invaluable, making it the preferred approach for complex projects (P1, P4–9, P11–12).

A Wilcoxon signed-rank test on exploratory counts showed a significant reduction with our system compared to the manual baseline ($p<.001$) as shown in Figure~\ref{fig:Study2Task1Result}-B. This indicates that our system reliably prunes irrelevant options, reducing the need for extensive search. Participants highlighted that this filtering substantially eased both cognitive and physical load (P12), allowing them to focus on a smaller set of high-quality choices (P11). They also noted that the system effectively eliminated options they would have dismissed anyway (P4, P5, P8), which streamlined the process and supported a more consistent decision-making approach.

Overall, the subjective ratings collected via Likert scales showed no significant differences between the baseline and our mapping module across all five categories (Ease of Decision, Decision Efficiency, Decision Confidence, Variety of Considerations, and Outcome Satisfaction) as shown in Figure~\ref{fig:Study2Task1Result}-C. This finding indicates that an automated, data-driven mapping approach can yield outcomes perceived as comparable in quality to those produced manually. Participants described the recommendations as ``appropriate'' and rarely in need of changes, noting that they often validated their own intentions (P4, P9) and were especially helpful when design goals were ambiguous (P9, P12). These results suggest that our system does not replace designer expertise but complements it, offering reliable guidance and effective HOI designs. This is particularly valuable when manual methods are difficult or designer experience is limited.

These results show that while the decision speed of \system{} was similar to manual design, it significantly reduced trial-and-error exploration and cognitive burden. Based on Likert-scale ratings, the outcomes were perceived as comparable to those created manually, indicating that the system captured many of the same considerations designers applied in their decisions. Designers were able to produce comparable HOI design results with less effort, underscoring the system's value as a supportive tool that enables mappings closer to expert-level outcomes. By reducing both cognitive and physical workload, \system{} not only broadens access to effective HOI authoring but also enhances overall productivity.

\subsubsection{Overall Results for Task 2}
\begin{figure*}[!t]
\centering
   \includegraphics[width=\textwidth]{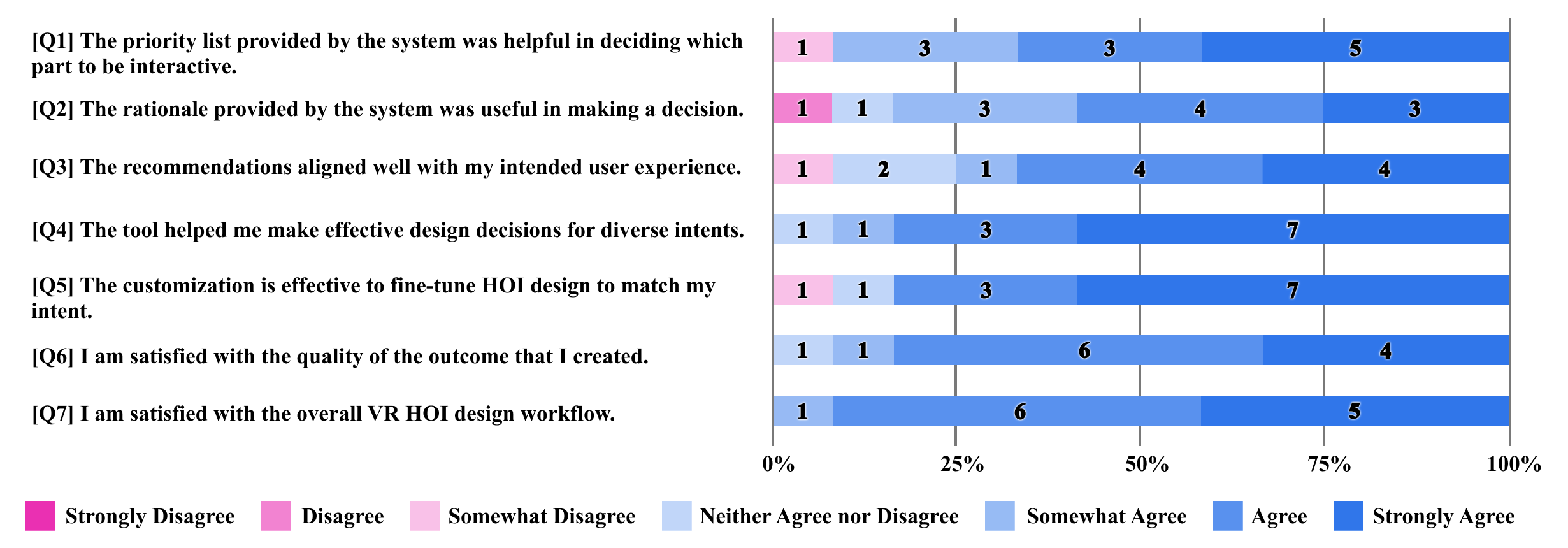}
   \hfil
\caption{Overall distribution of Likert-scale responses for the usability test. Results show the frequency of responses across all questionnaires.}
\Description{A stacked bar chart displaying Likert-scale responses for seven usability questions. The majority of participants responded with 'Agree' or 'Strongly Agree' for all items. Questions regarding satisfaction with the outcome (Q6) and overall workflow (Q7) show the highest positive responses.}
\label{fig:Study2Task2Likert}
\end{figure*}

Figure~\ref{fig:Study2Task2Likert} presents the overall results for Task 2. We asked Likert-scale responses related to our core modules to assess the usability of \system{}. 

\textbf{Interactive Part Selection~(Q1).}
Participants rated automated part selection as efficient and well-aligned with their intent ($Q1:M=5.81, SD=1.25$). They valued how the system filtered out less critical parts, especially for complex objects (P1, P3, P5, P8, P9, P12). P1 remarked, \textit{``The parts suggested at the start were exactly the ones I was thinking of. I don't need any change,''} and P3 added, \textit{``The initial recommendations were helpful and matched my intent; it also felt like the number of parts was chosen by importance.''} However, when designers had very specific goals, the system sometimes missed their intent, highlighting the need for hybrid use with manual refinement (P2, P7).

\textbf{HOI Mapping Module~(Q2-Q4).}
Feedback on the recommendation and rationale system was highly positive. The recommendations aligned well with participants' intended user experience ($Q3:M=5.54, SD=1.36$), validating both the system's effectiveness and their own design goals (P4, P8, P9, P11). The rationale further supported decision-making ($Q2:M=5.36, SD=1.43$), with some noting it was useful even when diverging from their initial intent, as it clarified the system's logic and built trust (P2, P5). Together, accurate recommendations and clear rationales enabled effective design decisions across diverse intents ($Q4:M=6.27, SD=1.00$). This was especially valuable when designers' goals were not fully formed, as the system guided them toward solutions they might not have considered (P9).

\textbf{Critical Role of Customization~(Q5).}
Customization was considered as essential ($Q5:M=6.09, SD=1.37$). Adjusting parameters like resistance or step angle was key to achieving realism and creative control, with participants noting it expanded the range of possible interactions (P2, P7, P9, P11). This feature enabled designers to tailor the system's output to their specific needs. As one participant noted, \textit{``Just by customizing the HOI design in this way, it seems like I could implement countless interactions''} (P2), suggesting that customization not only refined results but also unlocked broad design possibilities.

\textbf{Satisfaction~(Q6-7) \& SUS.}
Participants reported high satisfaction with both the outcomes ($Q6:M=6.0, SD=0.89$) and the overall workflow ($Q7:M=6.27, SD=0.64$). They valued intent input as a strong starting point (P4, P11), the efficiency of automated part recommendations (P3), and the ability to customize and test interactions in-situ, which enhanced both satisfaction and creativity (P5, P10). Even when suggestions did not perfectly align with their goals, participants described them as reasonable compromises that supported decision-making (P2, P5). Overall, the ability to refine and validate designs directly in VR made the workflow concise and efficient~(P10). The system also achieved a strong SUS score ($M=81.04, SD=13.75$), confirming its perceived usability.

\section{Discussion}

\begin{figure*}[t]
\centering
   \includegraphics[width=0.95\textwidth]{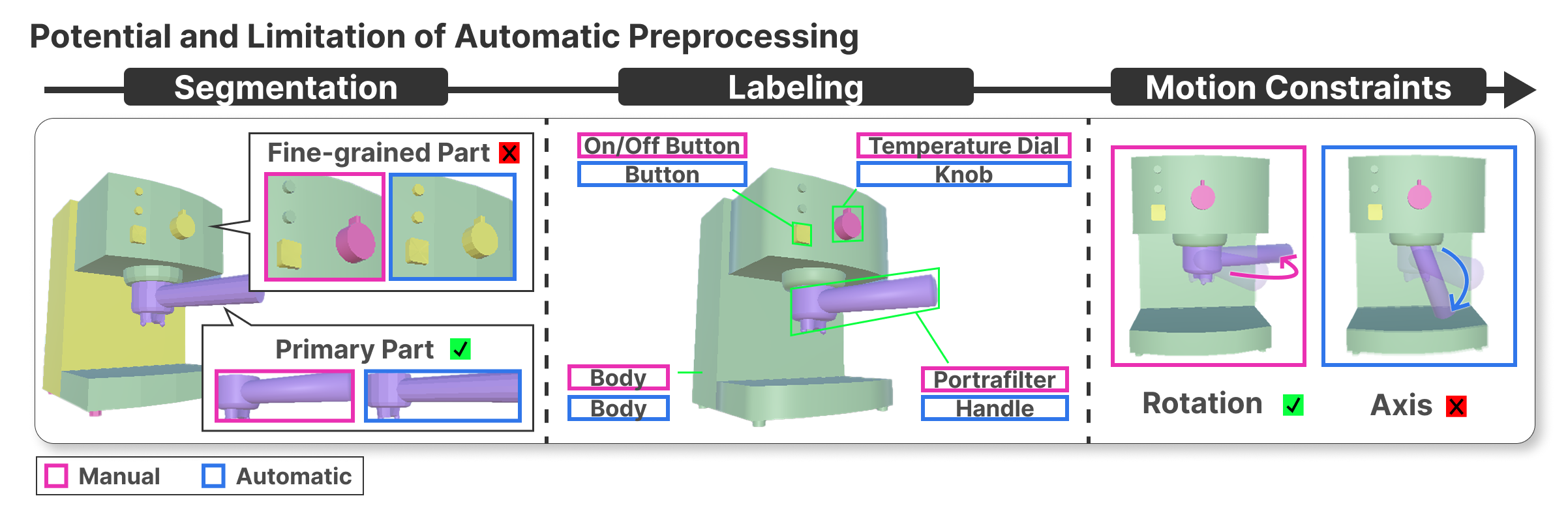}
   \hfil
\caption{Potential and limitation of SOTA automatic preprocessing. Segmentation using PartField~\cite{liu2025partfield} isolates primary components but fails to capture fine-grained functional parts. Labeling via VLM-based approaches~\cite{yang2024sampart3d, ArticulateAnymesh} accurately assigns basic names (e.g., button, handle) but struggles to identify specific functional semantics. Motion constraint inference using Articulate AnyMesh~\cite{ArticulateAnymesh} identifies plausible joint types (e.g., rotation) but fails to determine precise motion axes.}
\Description{A comparison diagram showing the potential and limitations of automatic preprocessing. In Segmentation, automatic methods identify large parts (primary) but fail on small details (fine-grained). In Labeling, automatic labels are generic (e.g., 'Knob') while manual ones are specific (e.g., 'Temperature Dial'). In Motion Constraints, automatic inference correctly identifies 'Rotation' but fails to determine the correct 'Axis', shown by the handle tilting incorrectly.}
\label{fig:Discussion}
\end{figure*}

Here, we reflect on the findings from the design and evaluation of \system{} and discuss lessons learned as well as directions for future AI-assisted HOI authoring tools.

\paragraph{Guiding Scaffold in Ambiguous Contexts.}
When design intent was vague or incomplete, participants highlighted that \system{} served as a valuable scaffold to support decision-making. Instead of forcing users to resolve uncertainty by themselves, the system’s capability of providing pros and cons alongside a recommended option set encouraged confidence. One participant explained, \textit{``Seeing the pros and cons and then a filtered set of options helped me make choices when the goal itself wasn’t very clear—like with general preference or easy \& intuitive''~(P12)}. Similarly, another remarked, \textit{``For something like general preference, the criteria felt ambiguous, so I tended to follow the recommendations''~(P9)}. These reflections suggest that \system{} was not only supporting decision-making, but also further providing design insights. Such scaffolding was especially seen as valuable in the early stages of interaction design. 

\paragraph{Supporting Novices and Experts Across the Design Spectrum.}
The main strength of \system{} lay in efficiency, serving as a bridge across different levels of expertise, supporting novices when choices feel difficult and relieving experts from repetitive effort. For novices, the recommendations simplified decision-making by narrowing down options and highlighting the most meaningful distinctions: \textit{``It reduced the factors I needed to consider, like whether to choose contact or gesture. It feels more efficient''~(P11)}. Similarly, P4 noted that the recommendations made comparisons easier, which simplified the process. This support often made it possible for beginners to decide with a confidence closer to expert decision-making: \textit{``After I got to know how HOIs work, I could first imagine the result of HOI and then test it. But with the recommended list already matching my expectations, I felt more confident. So I thought, especially for beginners, this makes it easier to decide like an expert''~(P7)}.

For experienced designers, the value of \system{} lay in reducing effort and accelerating iteration. Even with strong intuitions about HOI designs, the filtering process spared them from unnecessary comparisons and tedious repetitions. As one participant explained, \textit{``Even when I already knew how each HOI worked, it was convenient that the system excluded the less relevant ones, so I could pick more easily from the remaining set''~(P5)}. Another participant noted how this sped up iteration, allowing them to refine designs quickly without exhaustively checking every possibility~(P8). These findings suggest that recommendations can flexibly support designers at different stages: mitigating uncertainty and aiding decision-making for novices, while reducing the burden of exploration for experts. This dual role underscores the potential of AI-assisted authoring tools to adaptively enhance both learning and productivity across the spectrum of design expertise.

\paragraph{Encouraging Diverse Perspectives.}
Beyond efficiency, participants valued how \system{} encouraged them to reconsider their assumptions and explore alternatives they might not have otherwise considered. Unlike the baseline, where decisions were made in isolation, the system felt more like a collaborative partner—prompting reflection and sometimes compromise, similar to discussing design choices with teammates. One participant explained, \textit{``I initially thought \textbf{CM} would be better, but when the system recommended physics, I started wondering why. Then I realized that subtle effects like gravity could make it more appropriate. It made me pause and rethink, like a filter on my own decisions''~(P2)}. Participants also described how unexpected suggestions exposed them to options they had overlooked, \textit{``I thought gesture-based was the obvious choice, but the system pushed me to try something different, which turned out better than I expected''~(P8)}. Similarly, others appreciated how the pros-and-cons presentation revealed trade-offs they hadn’t noticed, \textit{``I hadn’t really considered contact vs. physics, but seeing the comparison made me realize the differences, and some options I thought were weak weren’t actually bad''~(P12)}. Together, these reflections highlight the system’s role in broadening design perspectives and viewpoints to enable designers to break from habitual choices and foster innovative outcomes.

\paragraph{Trade-Off between Design Flexibility and System Guidance.}
Our findings highlight the importance of flexibility in accommodating diverse designer working styles. Some participants with a strong and clear design vision preferred the manual baseline, as it allowed them to directly execute the choices they already had in mind without interference: \textit{``I already had a candidate in my head, so I found the baseline more straightforward''~(P1)}. In contrast, others valued the efficiency of automated pruning, especially when faced with numerous options to choose from. They noted that it helped reduce effort and mental load, \textit{``When parts increase, the system filtering is far more efficient—it saves me from trying everything one by one''~(P11)}.

A central design trade-off emerged around the system’s exclusion process. While a few participants felt constrained when their preferred option was filtered out (\textit{``Sometimes what I wanted wasn’t there, which was frustrating''~(P8)}), most appreciated the pruning, as it allowed them to focus on a smaller, high-quality set of choices (\textit{``It simplified what I had to consider and let me pay more attention to details''~(P11)}). These mixed reactions suggest that future systems should provide configurable levels of control. For instance, offering both a lightweight recommendation mode for quick decisions and a more exclusionary mode for curated and high-confidence guidance.

Another limitation participants noted was the difficulty of articulating design intent in free-text form. Abstract concepts such as ``intuitive'' or ``realistic'' were sometimes challenging to express, \textit{``It was hard to put intent into words directly''~(P3)}. Providing predefined intent presets or keywords could reduce this burden, particularly for novice designers or when the design goals are underspecified. Such presets could serve as starting points, which designers refine further, ensuring the system captures their intent while maintaining efficiency.

\section{Limitations and Future Work}

\paragraph{Handling Complex Interactions.}
In this work, we mainly evaluated interactions with prismatic and revolute joints (translation and rotation), as these serve as the majority of everyday articulated behaviors in VR interactions~\cite{lu2025dreamart, ArticulateAnymesh, chen2025freeart3d}. Still, we plan to further support unconventional interactions involving irregular geometries and multi-part coordination to broaden the system's scope to more complex articulated interactions. Specifically, for complex kinematics derived from irregular geometries, our current framework can be extended beyond 1-DoF movement by configuring motion constraints to accommodate multiple axes. For multi-part coordination, future work can integrate computational mechanism design~\cite{chen2024mpcmech, cheng2022exact} to model interdependent translations and rotations. This expansion would significantly strengthen the system’s applicability to a broader range of high-fidelity object types.

\paragraph{3D Object Preprocessing.}
\label{sec:ObjectPreprocessing}
Our framework relies on preprocessed objects~(segmented, labeled, and with defined motion constraints) to ensure interaction design workflows remain unaffected by artifacts arising from preprocessing. Still, we further investigated prospective directions for a scalable pipeline that encompasses automatic preprocessing. Recent advances in segmentation~\cite{liu2025partfield}, VLM-based labeling~\cite{yang2024sampart3d, ArticulateAnymesh}, and articulation inference~\cite{chen2025freeart3d, lu2025dreamart, ArticulateAnymesh} show promising capabilities. As shown in Figure~\ref{fig:Discussion}, our preliminary investigation showed that these methods can successfully generate useful initial proposals for structural and semantic mapping. However, challenges remain for future work, such as handling fine-grained parts in segmentation, identifying specific semantics in labeling, and determining precise parameters for motion constraints. Therefore, we envision a short-term hybrid approach where automatic systems provide a structural draft, allowing designers to efficiently validate and fine-tune the results through in-situ manual refinement. In the long term, the rapid evolution of 3D foundation models (e.g., SAM 3D~\cite{meta_sam3d_web_2025}) and advanced articulation inference have potential to resolve remaining issues, paving the way for a fully automated pipeline.
\paragraph{Extension to XR environments} While \system{} is currently implemented in VR, the core framework of adapting interaction mechanics to designers' intent remains critical for broader XR contexts. Since XR environments frequently utilize purely virtual elements, defining their interaction mechanics in accordance with user goals (e.g., precision vs. ease) remains a fundamental requirement. However, applying these mechanics in XR must also take into account real-world constraints. Therefore, future work will integrate our intent-driven logic with XR’s environmental awareness capabilities (e.g., spatial mapping) to create a system that respects physical boundaries while dynamically optimizing virtual behaviors.

\paragraph{Adaptability to Diverse Platforms.}
Our HOI mapping module is designed to be platform-agnostic, operating soley on designers' intent and object attributes rather than runtime-specific inputs. While our current execution layer is implemented using Meta's API, this layer can be adapted to any XR runtime (e.g., OpenXR, MRTK) that supports hand tracking. Future work will extend the execution layer into a modular framework, allowing the system to support diverse runtime setups.

\section{Conclusion}

We present \system{}, an AI-assisted authoring tool that reasons 3D objects as part-level interactive components in VR. \system{} integrates three core functions: \textit{Interactive Part Selection}, \textit{HOI Design Customization}, and \textit{HOI Design Mapping}. From our formative study, we derived five representative HOI designs and identified key metrics influencing their use. A data collection study then captured user preferences, informing the mapping module. The final user study that evaluates \system{} showed that although decision speed was comparable to manual workflows, \system{} significantly reduced trial-and-error exploration and cognitive load. Designers achieved outcomes perceived as similar in quality to manual results, with less effort. Participants emphasized the value of automated part selection, ranked recommendations with rationales, and customization for refining HOI mappings. In summary, \system{} demonstrates how AI can complement expertise, serving as both a practical assistant and a design partner for HOI authoring.

\begin{acks}
This work was supported by the Institute of Information \& Communications Technology Planning \& Evaluation(IITP) grant funded by the Korea government(MSIT) (IITP-RS-2025-02214780, Generative Haptics and Fine Response Inference for Flexible Tactile Interfaces)
\end{acks}

\bibliographystyle{ACM-Reference-Format}

    \bibliography{main}
\clearpage
\onecolumn
\appendix
\section{Object-Part Dataset}
\label{appendix:objectdataset} 
\begin{figure}[!h]
\centering
   \includegraphics[width=0.8\linewidth]{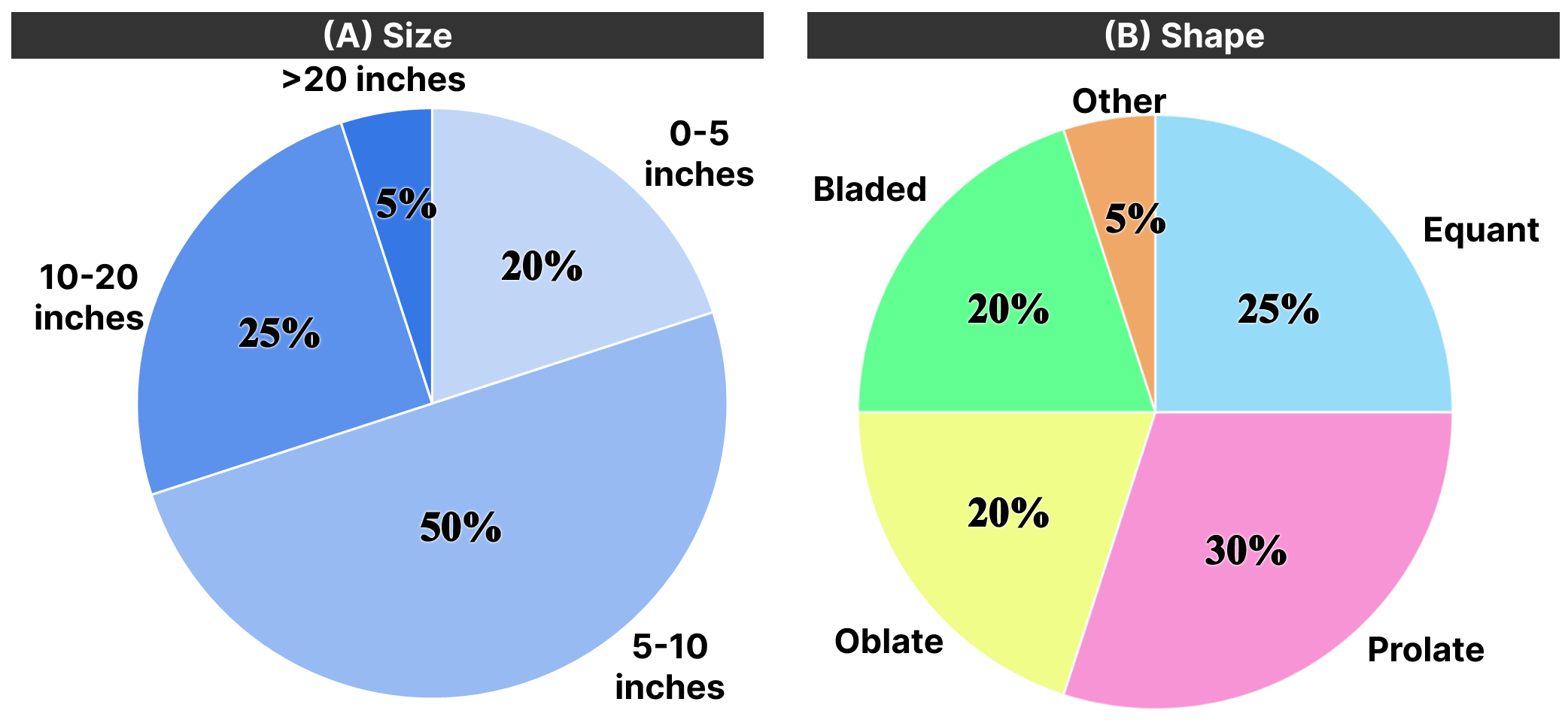}
   \hfil
\caption{The distribution of objects on the~(A)~size and~(B)~shape.}
\Description{Two pie charts showing object statistics. Panel A (Size) shows 50\% of objects are 5-10 inches, followed by 25\% for 10-20 inches. Panel B (Shape) shows a distribution of shapes: Prolate (30\%), Equant (25\%), Oblate (20\%), Bladed (20\%), and Other (5\%).}

\label{fig:objectDiversity}
\end{figure}

The object dataset for the formative study consisted of 20 everyday objects with articulated parts, selected to cover a broad range of affordances and motion types: laptop, scissors, drawer, syringe, cutter knife, stapler, door knob, faucet, spray bottle, dispenser, microwave, espresso machine, game controller, globe, mouse, kettle, camera, can, drill, and lightbulb. As shown in Figure~\ref{fig:objectDiversity}, these objects exhibit a balanced distribution across both size (0–5, 5–10, 10–20, and >20 inches) derived from Amazon\footnote{\url{https://www.amazon.com/}}, and shape categories~(equant, prolate, oblate, bladed, and other)~\cite{zingg1935beitrag, benn1993description}, ensuring diversity in geometry and interaction styles.

From the formative study, we adopted the \textit{game} context decomposition as the standard for object–part candidates, as it provided manageable part-level tasks aligned with our study goals. In contrast, \textit{training} contexts often produced overly detailed decompositions, while \textit{social} contexts typically resulted in minimal whole-object interactions.

\section{Tables for Likert Scale (\textit{Usability}, \textit{Realism})}
\label{appendix:Study1Task1Likert}
The following tables present key quantitative data from the Likert-scale responses in Section~\ref{sec:5DataCollection}, including Friedman test results with Kendall’s W and ranking significance. Methods are ordered by their mean scores (for Likert scales, higher values indicate better performance; for rank data, lower mean ranks indicate better performance). We then derived statistically informed groupings: methods not significantly different were grouped together (``=''), while significant differences formed a new, lower-ranked group (``>''). If the Friedman test was not significant, all methods were treated as tied.

For \textit{Usability}, we focused on Ease of use and Learnability as key factors. The results show largely consistent top tiers across the two, except for Part1 and Part5. To ensure flexibility, we adopted the data source that included more candidates in the top tier as the primary ranking measure for \textit{Usability}.

\begin{table}[!h]
\caption{Ease of Use results (Friedman test with Kendall’s W). Significance levels are indicated: * $p<0.05$, ** $p<0.01$, *** $p<0.005$, **** $p<0.001$).}
\begin{tabular}{c c l l}
\multicolumn{4}{l}{\small $n=20,\ k=5$} \\
\hline
Part & Kendall's W & Friedman\_sig & Ranking\_sig \\
\hline
1 & 0.3715 & p<0.001**** & \textbf{CA=GA}>CM=GM>PM \\
2 & 0.1883 & p=0.009** & \textbf{CM=CA}>GM=GA=PM \\
3 & 0.0948 & p=0.155 & \textbf{GM=CM=GA=CA=PM} \\
4 & 0.2135 & p=0.002*** & \textbf{CM=CA}>GM=GA>PM \\
5 & 0.1526 & p=0.050   & \textbf{CM=GM=CA}>GA=PM \\
6 & 0.1331 & p=0.050   & \textbf{CM=GM=CA=GA}>PM \\
7 & 0.2266 & p=0.002*** & \textbf{CM=CA=PM}>GM=GA \\
8 & 0.3385 & p<0.001**** & \textbf{CA=GA=GM=CM}>PM \\
9 & 0.0067 & p=0.969  & \textbf{GM=CM=CA=PM=GA} \\
10 & 0.1724 & p=0.010* & \textbf{CM=PM}>GM=CA>GA \\
11 & 0.2048 & p=0.003*** & \textbf{CM=CA=GA}>PM=GM \\
12 & 0.0364 & p=0.969   & \textbf{CM=PM=CA=GM=GA }\\
13 & 0.0872 & p=0.594   & \textbf{CA=GA=CM=PM=GM} \\
\hline
\end{tabular}
\Description{Ease of use results from Friedman tests with Kendall’s W, showing significant ranking differences among HOI designs.}
\end{table}

\begin{table}[!h]
\caption{Learnability results (Friedman test with Kendall’s W). Significance levels are indicated: * $p<0.05$, ** $p<0.01$, *** $p<0.005$, **** $p<0.001$).}
\begin{tabular}{c c l l}
\multicolumn{4}{l}{\small $n=20,\ k=5$} \\
\hline
Part & Kendall’s W & Friedman\_sig & Ranking\_sig \\
\hline
1 & 0.2153 & p=0.004*** & \textbf{CA=GA=CM}>GM=PM \\
2 & 0.1335 & p=0.049* & \textbf{CM=CA}>GM=GA=PM \\
3 & 0.0882 & p=0.432   & \textbf{GM=CM=GA=CA=PM} \\
4 & 0.1573 & p=0.019* & \textbf{CM=CA}>GM=GA=PM \\
5 & 0.1515 & p=0.035* & \textbf{CM=GM=CA=GA}>PM \\
6 & 0.1742 & p=0.008** & \textbf{CM=GM=CA=GA}>PM \\
7 & 0.2311 & p=0.001*** & \textbf{CA=CM=PM}>GM=GA \\
8 & 0.3224 & p<0.001**** & \textbf{CA=GM=GA=CM}>PM \\
9 & 0.0226 & p=0.771  & \textbf{PM=CM=GM=GA=CA} \\
10 & 0.1529 & p=0.020* & \textbf{CM=PM}>GM=CA=GA \\
11 & 0.1675 & p=0.010* & \textbf{CA=CM=GA}>PM=GM \\
12 & 0.0828 & p=0.771  & \textbf{PM=CM=CA=GM=GA} \\
13 & 0.0597 & p=0.771  & \textbf{CA=CM=GA=GM=PM }\\
\hline
\end{tabular}
\Description{Learnability results from Friedman tests with Kendall’s W, highlighting comparative performance across HOI designs.}
\end{table}

\begin{table}[!h]
\caption{Realism results (Friedman test with Kendall’s W). Significance levels are indicated: * $p<0.05$, ** $p<0.01$, *** $p<0.005$, **** $p<0.001$).}
\begin{tabular}{c c l l}
\multicolumn{4}{l}{\small $n=20,\ k=5$} \\
\hline
Part & Kendall’s W & Friedman\_sig & Ranking\_sig \\
\hline
1 & 0.3276 & p<0.001**** & \textbf{CM=GM}>PM=GA=CA \\
2 & 0.3334 & p<0.001**** & \textbf{CM=GM}>CA=PM=GA \\
3 & 0.4229 & p<0.001**** & \textbf{CM=GM=PM}>CA=GA \\
4 & 0.4031 & p<0.001**** & \textbf{CM}>GM=PM=CA>GA \\
5 & 0.2383 & p<0.001**** & \textbf{CM=GM}>PM=CA=GA \\
6 & 0.3613 & p<0.001**** & \textbf{CM=GM=PM}>CA=GA \\
7 & 0.1421 & p=0.073  & \textbf{CM=CA}>GM=PM>GA \\
8 & 0.2610 & p<0.001**** & \textbf{GM=CM}>CA=GA=PM \\
9 & 0.4006 & p<0.001**** & \textbf{GM=CM=PM}>GA=CA \\
10 & 0.5383 & p<0.001**** & \textbf{CM=PM}>GM=CA>GA \\
11 & 0.2354 & p<0.001**** & \textbf{CM=CA}>GM=PM=GA \\
12 & 0.3019 & p<0.001**** & \textbf{CM=GM=PM}>CA=GA \\
13 & 0.1678 & p=0.013* & \textbf{CM=GM=PM=GA}>CA \\
\hline
\end{tabular}
\Description{Realism results from Friedman tests with Kendall’s W, showing significant groupings of HOI designs.}
\end{table}

\clearpage

\section{System Prompts}
We provide the system prompts for each module in \system{} in this section. We prompted the GPT-4o with a temperature of 0.2 across all modules. 
 
\label{appendix:SystemPrompts}
\subsection{Object Analyzer}

\begin{lstlisting}%[label={lst:ObjectAnalzer}]
You are an extractor that outputs a simple, human-readable list of object parts.  
Task: Given an object schema (or description), return each part as one line in the format:

Rules:
- Interaction Type is the most typical action (e.g. Rotate, Click, Slide, Toggle).
- Affordances are the primary functions that parts provide to users.
- No extra text, just the lines.

### Input Format
Object: Object Name
Parts: Part Name1, Part Name2, ...

### Output Format
Note: The JSON output will be provided directly, without being enclosed in a code block.

[
  {
    "object": "objectName",
    "part": "partName",
    "interaction_type": "interactionType",
    "affordances": "affordance"
  },
  ...
]
\end{lstlisting}

\subsection{Part Prioritizer}
\begin{lstlisting}%[label={lst:InteractivePartSelection}]
You are an interaction authoring designer for VR.

Given an object with defined parts, a user intent, and constraints, decide:
1. The priority order of parts (skip the body).
2. A recommended initial complexity level based on intent/constraints.

The complexity level works as follows:
Level 1 = first (top) part only.
Level N (N >= 1) = top N parts.
Max level = number_of_parts.
The object body is implicit and excluded from level math.

### Input Format
{
  "intent": "<user_defined_intent>",
  "parts": [
    {"id": "part_id_1", "affordances": "part1_affordance"},
    {"id": "part_id_2", "affordances": "part2_affordance"},
    {"id": "part_id_3", "affordances": "part3_affordance"}
  ]
}

### Output Format
Note: The JSON output will be provided directly, without being enclosed in a code block.
{
  "priority_parts": ["part_id", "part_id", ...],
  "initial_level": selectedInteractionLevel,
  "rationale": "Explain in one or two sentences (max 150 characters) why this order and level match the intent."
}
\end{lstlisting}

\subsection{Metric Selector}
\label{appendix:metric_selector}
\textbf{Augmentation with Recommendation Metric }
We prompted the LLMs (Claude-Sonnet-4-20250514 and GPT-4o) with a temperature of 0.7 to augment the recommendation metric, and combined their outputs while removing redundancy.
\clearpage

\begin{table}[h!]
\centering
\begin{tabular}{|p{5cm}|p{10cm}|}
\hline
\textbf{System Prompt} & \textbf{Metric / Synonyms} \\
\hline
\multirow{4}{5cm}{Generate 15 synonyms or alternative terms for the given word in VR object interaction context. One word per line, maximum 1–2 words without numbering and explanations. All outputs must be adjectives only, without adverbs like ``highly'' or ``very''.  
} 
& \textbf{Realism}: Realistic, Lifelike, Authentic, Natural, Detailed, Immersive, Convincing, Credible, Believable, Vivid, True-to-life, Photorealistic, Faithful, Genuine \\ \cline{2-2}
& \textbf{Usability (Ease of Use, Learnability)}: Intuitive, Accessible, User-friendly, Navigable, Comprehensible, Effortless, Simple, Streamlined, Clear, Approachable, Responsive, Comfortable   \\ \cline{2-2}
& \textbf{Challenge}: Demanding, Complex, Challenging, Intense, Skillful, Rewarding, Testing, Strenuous, Satisfying, Motivating, Intricate, Mastery-focused, Accomplishment-driven, Stimulating, Engaging \\
\hline 
\end{tabular}
\end{table}

\begin{lstlisting}%[label={lst:MetricSelector}]
You are a metric selector for evaluating Hand-Object Interaction (HOI) techniques. 

### Goal: Read the intent and decide which evaluation metrics should be prioritized among: realism, usability, preference, efficiency, challenge.

### Output Format
Note: The JSON output will be provided directly, without being enclosed in a code block.
You must return the result in the following strict format:
[
  {"part": part, "metric": metric, "reason": reason},
  {"part": part, "metric": metric, "reason": reason},
]

### Input Format: intent: [interactive parts] and [user intent text] 

### Strict Keyword-Based Decision Rules: 

**REALISM**: Select ONLY if intent contains explicit realism keywords: Realistic, Lifelike, Authentic, Natural, Detailed, Immersive, Convincing, Credible, Believable, Vivid, True-to-life, Photorealistic, Faithful, Genuine

**USABILITY**: Select ONLY if intent contains usability keywords: Intuitive, Accessible, User-friendly, Navigable, Comprehensible, Effortless, Simple, Streamlined, Clear, Approachable, Responsive, Comfortable

**EFFICIENCY**: Select ONLY if the intent emphasizes quantitative performance of the interaction. 
This includes references to task completion time, execution accuracy, error/reversal counts, or performance-related terms such as speed, responsiveness, low latency, and smooth task flow.

**CHALLENGE**: Select ONLY if intent contains challenge keywords: Demanding, Complex, Challenging, Intense, Skillful, Rewarding, Testing, Strenuous, Satisfying, Motivating, Intricate, Mastery-focused, Accomplishment-driven, Stimulating, Engaging

**PREFERENCE**: Select if: - lacks clear keywords for the above three metrics - OR expresses subjective desires, or individual taste 

### Priority Order: 
1. Check for explicit REALISM keywords first 
2. Check for explicit USABILITY keywords second 
3. Check for explicit EFFICIENCY keywords third 
4. Check for explicit CHALLENGE keywords fourth
5. Default to PREFERENCE for all other cases 

### Examples:
- "I want realistic physics" - realism (contains "realistic") 
- "Make it easy for beginners" - usability (contains "easy", "beginners") 
- "Need fast response times" - efficiency (contains "fast") 
- "I want to master a difficult skill" - challenge (contains "master", "difficult")
- "I want something that feels right for my workflow" - preference (semantic: personal satisfaction)
\end{lstlisting}

\subsection{Matching Similar Part}

\begin{lstlisting}%[label={lst:MatchingSimilarPart}]
You are an expert interaction designer specializing in hand-object interactions in virtual environments. Your task is to identify the most similar object-parts from a database, based on a newly described part.

### Matching criteria:
1. Gesture/Interaction Type: Match the required hand pose and motion (e.g., pinch, grab, twist, slide, push, pull, tap, press).
2. Physical Shape & Size: Classify the required movement scale (e.g., small movements like finger or wrist control vs. large movements like arm control). Note the presence of a graspable handle.
3. Motion Constraints: Determine if the motion is rotational or linear, and if it's bounded or unbounded.
4. Control Granularity: Identify if the action is a continuous adjustment (fine control, variable range) or a discrete/binary action (on/off, open/close, stepwise).

### Input Format
You will receive a list of object parts in the format:
[ObjectName-PartName-Interaction Type]

### Output Format
Note: The JSON output will be provided directly, without being enclosed in a code block.
You must return the result in the following strict format:
[
  {"part": "PartName", "id": MatchedPartId , "matchedPart" : "MatchedPartName"}, ...
]

### Rules
- Return exactly **1 best match** per input part.
- Use only the provided **Given Object List** for matching.
- Each match must return its original number (id) from the list along with the name.
- Do not include explanations or any extra text.
- The "part" field must include only the PartName from the input (e.g., from "Padlock-Dial-Rotate", use only "Dial").
- The number of output objects must equal the number of input parts. Never return all items from the Given Object List.

### Given Object List
1. Laptop-Hinge - Rotating hinge for opening/closing the screen 
2. Scissors-Handle - Looped finger grips for squeezing the blades 
3. CutterKnife-BladeSlider - Thumb slider that adjusts blade length
4. Stapler - Press-down mechanism that drives a staple 
5. Doorknob-Lever - Lever rotated to unlatch/open the door
6. SprayBottle-Trigger - Finger-squeezed trigger that releases mist
7. PumpBottle-PumpHead - Press-down pump head that dispenses liquid 
8. Microwave-Door - Hinged door pulled to access the chamber 
9. Microwave-Dial- Flat Rotatable-button for function control
10. Globe-Sphere - Rotatable sphere for exploration/navigation 
11. Camera-ShutterButton - Index-finger button for capturing images 
12. Padlock-Combination Dial - Rotating numbered dial for combination input 
13. Padlock-Shackle - U-shaped bar that lifts when unlocked
\end{lstlisting}

\subsection{HOI Mapper Based on Ranking-based Decision}

\begin{lstlisting}%[label={lst:RankingHOIMapper}]
You are an assistant that recommends the most appropriate Hand-Object Interaction (HOI) technique for a given 3D object part in VR. 

### Input Format
1. A set of candidate HOIs (already statistically significant).
2. Raw participant comments describing pros and cons of each HOI.
3. A user intent describing how the object should be used in VR.

### Your task:
1. Rank the candidate HOIs from most to least appropriate based on the user's intent.
2. For each ranked HOI, justify its position by summarizing the most relevant participant comments into keywords for pros and cons.
**Do not include participant IDs (e.g., P1, P2) and words "participant(s)".** 
**Instead, present the reasoning directly as neutral evidence (e.g., "This technique is convenient and requires less effort...").**
3. Use only the provided candidates (e.g., PM, GM, GA). Never recommend excluded HOIs.
4. Assign higher weight to evidence directly linked to the intent's priority.

There are five available HOI techniques:
1. Physics-based Manipulation (PM)
The user's hand collider directly collides with the object's collider.
The object moves according to game engine's physical interaction.

2. Gesture-based Manipulation (GM)
The user performs a gesture near the object (e.g., grab, pinch).
The object then follows the user's hand movement.

3. Gesture-based Animation (GA)
The user performs a gesture near the object (e.g., grab, pinch).
Instead of following the hand, the object moves automatically along a predefined animation created by the developer.

4. Contact-based Manipulation (CM) 
When the user's hand approaches the object, the object snaps and follows the hand's movement. No explicit gesture is required.

5. Contact-based Animation (CA)
When the user's hand approaches the object, the object automatically executes a predefined animation created by the developer. No explicit gesture is required.

Provide the result strictly in the JSON format below.

### Output Format
Note: The JSON output will be provided directly, without being enclosed in a code block.
[
  {
    "rank": 1,
    "choice": "PM|GM|GA|PM|PA",
    "rationale": "A 1-2 sentence rationale explaining why this HOI is the top choice. (max 150 chars)",
    "keywords": {
      "pros": ["Pro keyword 1", "Pro keyword 2"],
      "cons": ["Con keyword 1", "Con keyword 2"]
    }
  },
  {
    "rank": 2,
    "choice": "PM|GM|GA|PM|PA",
    "rationale": "A 1-2 sentence rationale explaining why this HOI is the second choice. (max 150 chars)",
    "keywords": {
      "pros": ["Pro keyword 1", "Pro keyword 2"],
      "cons": ["Con keyword 1", "Con keyword 2"]
    }
  }
]
\end{lstlisting}
\subsection{HOI Mapper Based on Binary Decision}
\begin{lstlisting}%[label={lst:BinaryHOIMapper}]
You are an assistant that recommends the most appropriate Hand-Object Interaction (HOI) technique for a given 3D object part in VR. 

There are three available HOI techniques:
1. Physics-based Manipulation (PM)
The user's hand collider directly collides with the object's collider.
The object moves according to game engine's physical interaction.

2. Gesture-based Manipulation (GM)
The user performs a gesture near the object (e.g., grab, pinch).
The object then follows the user's hand movement.

3. Contact-based Manipulation (CM) 
When the user's hand approaches the object, the object snaps and follows the hand's movement. No explicit gesture is required.


### Decision Rules:
1. If the primary metric is Efficiency:
If the intent focuses on precision (fine-tuning, delicate actions) and control, select GM.
If the intent focuses on speed (fast actions, minimal effort) and is easy, select CM.

2. If the primary metric is Challenge:
If the sub-intent is focused on mastery and skill development, select GM.
If the sub-intent is focused on realistic difficulty and natural resistance, select PM.

### Input Format:
primary_metric: [Efficiency | Challenge] intent: [user's intent text]

### Output Format
Note: The JSON output will be provided directly, without being enclosed in a code block.
[
  {
    "rank": 1,
    "choice": "PM|GM|CM",
    "rationale": "A 1-2 sentence rationale explaining why this HOI is the top choice.",
    "keywords": {
      "pros": ["Pro keyword 1", "Pro keyword 2"],
      "cons": ["Con keyword 1", "Con keyword 2"]
    }
  },
  {
    "rank": 2,
    "choice": "PM|GM|CM",
    "rationale": "A 1-2 sentence rationale explaining why this HOI is the second choice.",
    "keywords": {
      "pros": ["Pro keyword 1", "Pro keyword 2"],
      "cons": ["Con keyword 1", "Con keyword 2"]
    }
  }
]
\end{lstlisting}

\end{document}